\documentclass[final,onefignum,onetabnum]{siamonline190516}

\usepackage{mathtools}
\usepackage{booktabs}
\usepackage[english]{babel} 
\usepackage[protrusion=true,expansion=true]{microtype} 
\usepackage{braket,amsfonts}
\usepackage{ amssymb }
\usepackage{enumitem}
\usepackage{graphicx}
\usepackage{array}
\usepackage[toc,page]{appendix}

\usepackage{booktabs} 
\usepackage{setspace}
\usepackage{relsize}

\usepackage{microtype} 
\usepackage{tabularx}
\usepackage{footnote}
\usepackage{bbm}
\usepackage{multirow}
\usepackage{tikz}
\usetikzlibrary{shapes,arrows}

\newsiamthm{claim}{Claim}
\newsiamremark{remark}{Remark}
\newcolumntype{C}[1]{>{\centering\let\newline\\\arraybackslash\hspace{0pt}}m{#1}}

\DeclareMathOperator*{\argmin}{\arg\!\min}

\usepackage{color}

\newcommand{\cmtS}[1]{{\color{red}{(Simon: #1)}}}

\newcommand{\cmtI}[1]{{\color{magenta}{(Irene: #1)}}}

\newcommand{\bi}{\begin{itemize}}
\newcommand{\ei}{\end{itemize}}

\usepackage[caption=false]{subfig}
\usepackage{algorithm}
\usepackage{algpseudocode}

\usepackage{graphicx,epstopdf}

\Crefname{ALC@unique}{Line}{Lines}

\usepackage{amsopn}

\usepackage{xspace}
\usepackage{bold-extra}
\usepackage[most]{tcolorbox}

\colorlet{texcscolor}{blue!50!black}
\colorlet{texemcolor}{red!70!black}
\colorlet{texpreamble}{red!70!black}
\colorlet{codebackground}{black!25!white!25}

\lstdefinestyle{siamlatex}{%
  style=tcblatex,
  texcsstyle=*\color{texcscolor},
  texcsstyle=[2]\color{texemcolor},
  keywordstyle=[2]\color{texemcolor},
  moretexcs={cref,Cref,maketitle,mathcal,text,headers,email,url},
}

\tcbset{%
  colframe=black!75!white!75,
  coltitle=white,
  colback=codebackground, 
  colbacklower=white, 
  fonttitle=\bfseries,
  arc=0pt,outer arc=0pt,
  top=1pt,bottom=1pt,left=1mm,right=1mm,middle=1mm,boxsep=1mm,
  leftrule=0.3mm,rightrule=0.3mm,toprule=0.3mm,bottomrule=0.3mm,
  listing options={style=siamlatex}
}

\newtcblisting[use counter=example]{example}[2][]{%
  title={Example~\thetcbcounter: #2},#1}

\newtcbinputlisting[use counter=example]{\examplefile}[3][]{%
  title={Example~\thetcbcounter: #2},listing file={#3},#1}

\DeclareTotalTCBox{\code}{ v O{} }
{ 
  fontupper=\ttfamily\color{black},
  nobeforeafter,
  tcbox raise base,
  colback=codebackground,colframe=white,
  top=0pt,bottom=0pt,left=0mm,right=0mm,
  leftrule=0pt,rightrule=0pt,toprule=0mm,bottomrule=0mm,
  boxsep=0.5mm,
  #2}{#1}

\patchcmd\newpage{\vfil}{}{}{}
\begin{tcbverbatimwrite}{tmp_\jobname_header.tex}
\title{Stacking designs: designing multi-fidelity computer experiments with target predictive accuracy \thanks{Submitted to the editors on Oct. 31, 2022.
\funding{CS gratefully acknowledges funding from NSF DMS 2113407. YJ and SM are funded by NSF CSSI Frameworks 2004571, NSF DMS 2210729, NSF DMS 2316012 and DE-SC0024477. TT is funded by the Statistical and Applied Mathematical Sciences Institute.}}}

\author
{Chih-Li Sung \thanks{Department of Statistics and Probability, Michigan State University (\email{sungchih@msu.edu}).}
\and
Yi (Irene) Ji \thanks{Department of Statistical Science, Duke University (\email{yi.ji@duke.edu}).
} 
\and
Simon Mak \thanks{Department of Statistical Science, Duke University (\email{sm769@duke.edu}).}
\and
Wenjia Wang \thanks{Data Science and Analytics Thrust, Information Hub, Hong Kong University of Science and Technology  (Guangzhou) (\email{wenjiawang@ust.hk}).} 
\and
Tao Tang \thanks{Department of Mathematics, Duke University (\email{tao.tang250@duke.edu}).} 
}

\end{tcbverbatimwrite}
\title{Stacking designs: designing multi-fidelity computer experiments with target predictive accuracy \thanks{Submitted to the editors on Oct. 31, 2022.
\funding{CS gratefully acknowledges funding from NSF DMS 2113407. YJ and SM are funded by NSF CSSI Frameworks 2004571, NSF DMS 2210729, NSF DMS 2316012 and DE-SC0024477. TT is funded by the Statistical and Applied Mathematical Sciences Institute.}}}

\author
{Chih-Li Sung \thanks{Department of Statistics and Probability, Michigan State University (\email{sungchih@msu.edu}).}
\and
Yi (Irene) Ji \thanks{Department of Statistical Science, Duke University (\email{yi.ji@duke.edu}).
}
\and
Simon Mak \thanks{Department of Statistical Science, Duke University (\email{sm769@duke.edu}).}
\and
Wenjia Wang \thanks{Data Science and Analytics Thrust, Information Hub, Hong Kong University of Science and Technology  (Guangzhou) (\email{wenjiawang@ust.hk}).}
\and
Tao Tang \thanks{Department of Mathematics, Duke University (\email{tao.tang250@duke.edu}).}
}

\ifpdf
\hypersetup{ pdftitle={Guide to Using  SIAM'S \LaTeX\ Style} }
\fi

\begin{document}

\maketitle

\begin{tcbverbatimwrite}{tmp_\jobname_abstract.tex}
\begin{abstract}
In an era where scientific experiments can be very costly, multi-fidelity emulators provide a useful tool for cost-efficient predictive scientific computing. For scientific applications, the experimenter is often limited by a tight computational budget, and thus wishes to (i) maximize predictive power of the multi-fidelity emulator via a careful design of experiments, and (ii) ensure this model achieves a desired error tolerance with some notion of confidence. Existing design methods, however, do not jointly tackle objectives (i) and (ii). We propose a novel stacking design approach that addresses both goals. A  multi-level reproducing kernel Hilbert space (RKHS) interpolator is first introduced to build the emulator, under which our stacking design provides a sequential approach for designing multi-fidelity runs such that a desired prediction error of $\epsilon > 0$ is met under regularity assumptions. We then prove a novel cost complexity theorem that, under this multi-level interpolator, establishes a bound on the computation cost (for training data simulation) needed to achieve a prediction bound of $\epsilon$. This result provides novel insights on conditions under which the proposed multi-fidelity approach improves upon a conventional RKHS interpolator which relies on a single fidelity level. Finally, we demonstrate the effectiveness of stacking designs in a suite of simulation experiments and an application to finite element analysis.

\end{abstract}

\begin{keywords} Computer Experiments, Experimental Design, Finite Element Analysis, RKHS interpolator, Multi-level Modeling, Uncertainty Quantification.
\end{keywords}

\begin{AMS}
  00A20 
\end{AMS}
\end{tcbverbatimwrite}
\begin{abstract}
In an era where scientific experiments can be very costly, multi-fidelity emulators provide a useful tool for cost-efficient predictive scientific computing. For scientific applications, the experimenter is often limited by a tight computational budget, and thus wishes to (i) maximize predictive power of the multi-fidelity emulator via a careful design of experiments, and (ii) ensure this model achieves a desired error tolerance with some notion of confidence. Existing design methods, however, do not jointly tackle objectives (i) and (ii). We propose a novel stacking design approach that addresses both goals. A  multi-level reproducing kernel Hilbert space (RKHS) interpolator is first introduced to build the emulator, under which our stacking design provides a sequential approach for designing multi-fidelity runs such that a desired prediction error of $\epsilon > 0$ is met under regularity assumptions. We then prove a novel cost complexity theorem that, under this multi-level interpolator, establishes a bound on the computation cost (for training data simulation) needed to achieve a prediction bound of $\epsilon$. This result provides novel insights on conditions under which the proposed multi-fidelity approach improves upon a conventional RKHS interpolator which relies on a single fidelity level. Finally, we demonstrate the effectiveness of stacking designs in a suite of simulation experiments and an application to finite element analysis.

\end{abstract}

\begin{keywords} Computer Experiments, Experimental Design, Finite Element Analysis, RKHS interpolator, Multi-level Modeling, Uncertainty Quantification.
\end{keywords}

\begin{AMS}
  00A20
\end{AMS}

\section{Introduction}
With recent developments in scientific computing and mathematical modeling, computer experiments have now become an essential tool in solving many scientific and engineering problems. These experiments, which typically solve complex mathematical models representing reality, are useful in applications which are prohibitively expensive or infeasible for direct experimentation. Such virtual experiments have now been successfully applied in a broad range of problems, from nuclear physics \cite{everett2021multisystem} to rocket design \cite{mak2017efficient}. As the science becomes more sophisticated, however, such simulations can become prohibitively costly for parameter space exploration. A popular solution is \textit{emulation} \cite{santner2003design}, which makes use of a carefully designed training set from the simulator to build an efficient predictive model that \textit{emulates} the expensive computer code. Popular emulator models include the Gaussian process (GP) model or reproducing kernel Hilbert space (RKHS) interpolators \cite{sacks1989design,currin1991bayesian,gramacy2020surrogates,haaland2011accurate}, neural networks \cite{tripathy2018deep,myren2021comparison} and polynomial chaos methods \cite{xiu2010numerical}, all of which have demonstrated successes in various areas of application. 

For full-scale complex scientific systems, however, it is often the case that the training data needed to train an accurate emulator model can be prohibitively expensive to generate from the simulator. One way to address this is via \textit{multi-fidelity emulation}, which supplements the costly high-fidelity (or high-accuracy) simulation dataset with less expensive lower-fidelity (or lower-accuracy) approximations for fitting the emulator model. The idea is that, by leveraging useful information from cheaper lower-fidelity simulations to enhance predictions for the high-fidelity model, an accurate multi-fidelity emulator can be trained with fewer high-fidelity runs and thus lower simulation costs.
The usefulness of this multi-fidelity emulation framework has led to much work in recent years. A popular framework is the Kennedy-O'Hagan (KO) model \cite{kennedy2000predicting}, which models a sequence of computer simulations from lowest to highest fidelity using a sequence of GP models linked by a linear autoregressive framework. Recent developments on the KO model include \cite{qian2008bayesian,le2013bayesian,le2014recursive,le2015cokriging,Perdikaris_2017,ji2022graphical} (among many others), which investigated modeling strategies for efficient posterior prediction and Bayesian uncertainty quantification. For multi-fidelity simulators controlled by a single mesh parameter (e.g., mesh density in finite element analysis), \cite{tuo2014surrogate} proposed a non-stationary GP model which leverages data at different mesh densities to predict the highest-fidelity simulation at the finest mesh. This model has been further developed for conglomerate multi-fidelity emulation \cite{ji2022multi} and graphical multi-fidelity emulation \cite{ji2022graphical}.


Despite this body of work, there remains important unresolved needs, particularly on the \textit{design} of such multi-fidelity experiments for cost-efficient emulation. In modern scientific computing problems, the experimenter is often limited by a tight computational budget dictated by available computing resources for a project; see, e.g., \cite{cao2021determining}. Given such constraints, one thus wishes to (i) \textit{maximize} predictive power of the multi-fidelity emulator via a careful design of experiments, and (ii) ensure the resulting emulator achieves a desired prediction error bound with some notion of \textit{confidence}. Despite its importance, however, there is little work to our knowledge on design methods for multi-fidelity emulators which jointly tackles objectives (i) and (ii). Recent works have addressed objective (i) in various ways. For instance, \cite{le2015cokriging} propose one-sample-at-a-time and batch sequential designs using fast cross-validation techniques for multi-fidelity simulations. \cite{stroh2022sequential} suggest a sequential design that maximizes uncertainty reduction while considering the simulation cost. \cite{ehara2021} introduce a sequential design strategy that maximizes the \textit{mutual information} for the next sampling location and minimizes the theoretical error bound to select the most effective fidelity level. However, these works do not fully address objective (ii). Hence, we propose a novel \textit{stacking design} framework which aims to jointly address (i) and (ii).



In what follows, we let $f_l(x)$ denote the scalar (deterministic) simulation output of the computer code with input parameters $x\in\Omega\subseteq\mathbb{R}^d$ and at {fidelity} level $l$. In finite element analysis (FEA), this fidelity level may reflect the underlying mesh density of the numerical simulator. We further suppose that, as fidelity level $l$ increases, the simulated output $f_l(x)$ approaches a limiting solution of $f_\infty(x)$, the desired ``exact'' solution of the simulator. In practice, this exact solution often cannot be simulated numerically; for example, the limiting solution for finite element analysis (i.e., at an infinitely dense mesh density) often cannot be computed numerically. Our proposed method makes use of a multi-level RKHS interpolator, which leverages simulated data of multiple fidelities to train  an emulator for predicting the desired limiting solution $f_\infty(x)$.

\begin{figure}[!t]
    \centering
    \includegraphics[width=\textwidth]{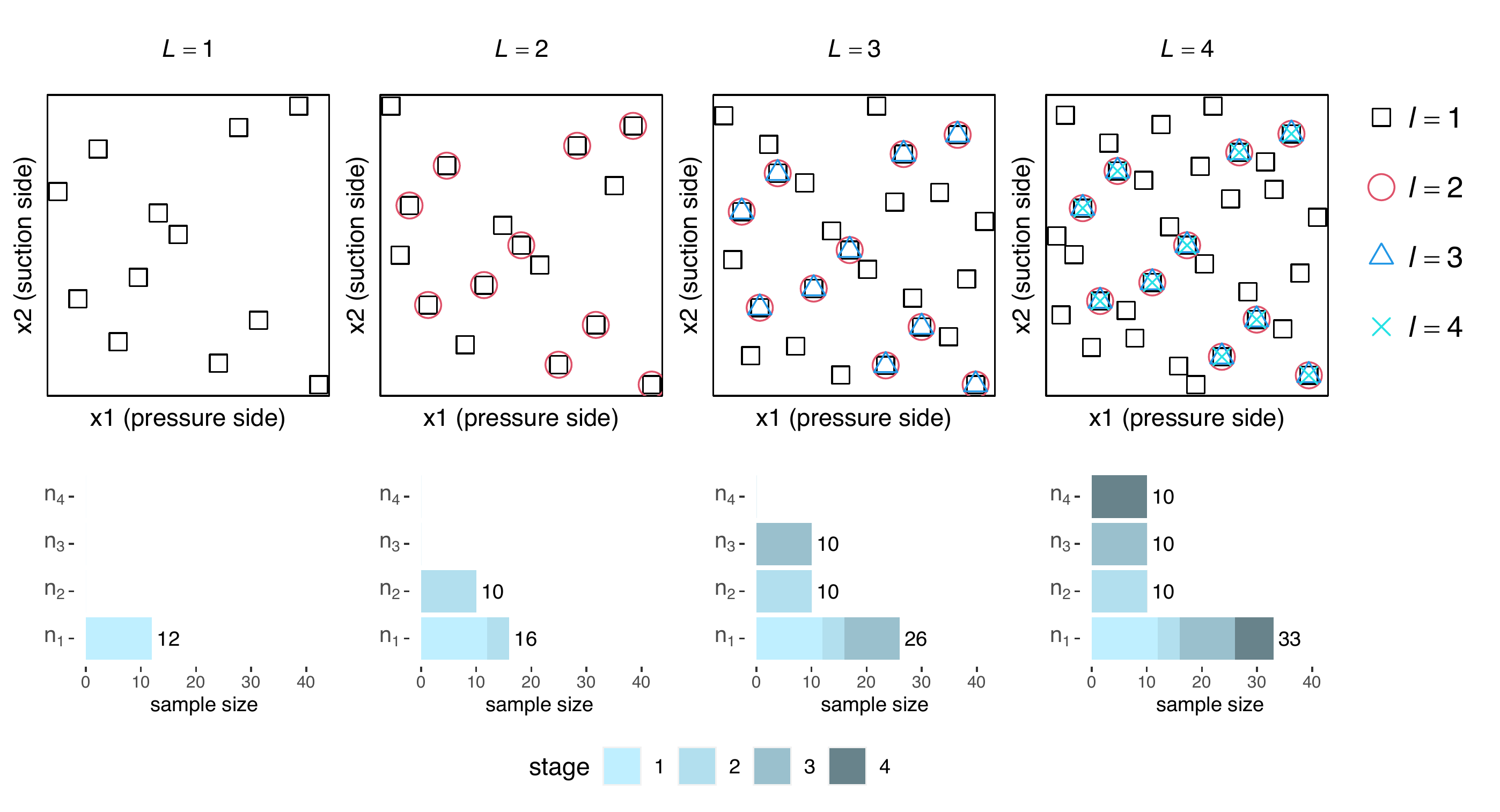}
    \caption{Visualizing the stacking behavior of the proposed stacking designs for a jet engine turbine blade application investigated later. (Top) The proposed stacking designs at $L=4$ fidelity levels, over four batch sequential design stages (from left to right). (Bottom) The corresponding sample sizes at the $L=4$ fidelity levels, over four batch sequential design stages (from left to right).}
    \label{fig:blade_design}
\end{figure}

With the multi-level interpolator, the proposed stacking design aims to carefully choose the multi-fidelity experimental runs, to target a desired prediction bound of $\epsilon > 0$ between the multi-level interpolator and the desired solution $f_\infty(\cdot)$. This is achieved by iterating the following sampling steps. First, for a fixed number of sampled fidelity levels $L$, we show that the multi-level interpolator provides an easy-to-evaluate expression for allocating sample sizes over each fidelity level. After performing runs with these sample sizes in a space-filling fashion, we then derive a novel stopping rule for deciding, under regularity conditions, whether the number of levels $L$ is sufficiently large for achieving the desired error bound $\epsilon$. If $L$ is not large enough, we then increment $L$ and repeat the above design step. The resulting batch sequential design procedure creates a ``stacking'' effect, where design points are stacked on at each fidelity level until the stopping rule is satisfied, at which point the desired prediction bound $\epsilon$ is satisfied under regularity conditions. This stacking behavior is visualized in the bottom of Figure \ref{fig:blade_design}, where we see the ``stacking'' of sample sizes at each fidelity stage as the batch sequential design progresses. The proposed stacking designs are inspired by a similar sequential approach for multi-level Monte Carlo (MLMC; see \cite{giles2008multilevel}), which aims to provide cost-efficient error control for multi-level Monte Carlo simulations. We then demonstrate the effectiveness of the proposed stacking designs (in terms of cost efficiency and error guarantees) in a suite of simulation experiments and an application to finite element analysis.

A key novelty of our work is a new \textit{cost complexity theorem} which, under the multi-level interpolator, establishes a bound on the required computation cost (for training data generation) needed to ensure the desired prediction bound $\epsilon$. Such a result sheds useful insight on when multi-fidelity emulation may be most (or least) effective given a computation cost budget. As a corollary, we then show that the presented multi-fidelity approach yields provably improved predictions over a conventional single-fidelity RKHS interpolator, under intuitive conditions on the error decay and cost complexity of the multi-fidelity simulator. These results are again inspired by existing work in MLMC for characterizing the cost complexity of multi-level Monte Carlo estimators. To our knowledge, there has been little work on extending such results for characterizing cost complexity of multi-fidelity computer experiments; we aim to address this gap here.

The paper is organized as follows. Section \ref{sec:model} introduces the multi-level RKHS interpolator. Section \ref{sec:nofidelity} presents the proposed stacking designs in several steps. Section \ref{sec:complexitytheorem} discusses a cost complexity theorem for the  multi-level interpolator. Section \ref{sec:numericstudy} investigates the effectiveness of stacking designs via a suite of simulation studies and an application to FEA. Section \ref{sec:discussion} concludes the paper. Proofs and code for reproducing numerical results are provided in Supplemental Materials.

\section{Multi-Level RKHS Interpolator}\label{sec:model}



We first introduce a multi-level interpolator, which we leverage later for our stacking designs. Again, let $f_l(x)$ denote the scalar simulation output of the computer code, with input parameters $x\in\Omega\subseteq\mathbb{R}^d$ and at fidelity level $l$. In what follows, we assume that $L$ distinct fidelity levels have been sampled for training data, where a larger fidelity level indicates a higher fidelity (or higher accuracy) simulator with higher computational costs per run.

The goal then is to construct an efficient surrogate model, with uncertainty quantification, for the highest-fidelity (and thus most expensive) simulation code $f_L(x)$. Suppose, for $l$-th fidelity level, simulations are performed at the design points $\mathcal{X}_l = \{x_i^{[l]}\}_{i=1}^{n_l}$, where the sample size $n_l$ varies for different fidelity levels $l$. This yields the simulation outputs $f_l|_{\mathcal{X}_l} = (f_l(x))_{x \in \mathcal{X}_l}$, where $f|_\mathcal{X}$ denotes the vector of outputs for $f(x)$ at design points $x \in\mathcal{X}$. For this multi-level emulator, we further assume that the designs $\mathcal{X}_l$ are sequentially nested, i.e.,
\begin{equation}
\mathcal{X}_L\subseteq \mathcal{X}_{L-1}\subseteq\cdots\subseteq \mathcal{X}_1\subseteq \Omega,
\label{eq:subset}
\end{equation}
In other words, design points run for a higher fidelity simulator will be contained within the design points run for a lower fidelity simulator.

With this, the multi-level interpolator is constructed as follows. Note that the high-fidelity response surface $f_L$ can be decomposed as $f_L=\sum^L_{l=1}(f_l-f_{l-1})$, where $f_0 \equiv 0$, and $(f_l-f_{l-1})$ can be viewed as the \textit{discrepancy} between the $(l-1)$-th and $l$-th code, capturing \textit{refinements} in the response surface as fidelity increases. Suppose that $\Phi_l(x,x')$ is a positive-definite kernel function used for interpolating the refinement $(f_l-f_{l-1})|_{\mathcal{X}_l}$, then the RKHS interpolator of $(f_l-f_{l-1})|_{\mathcal{X}_l}$ has the simple form \cite{wendland2004scattered,haaland2011accurate}:
$$
\mathcal{P}_l(x)=\sum^{n_l}_{i=1}\alpha^l_i\Phi_l(x,x^{[l]}_i),
$$
where $(\alpha^l_1,\ldots,\alpha^l_{n_i})^T=\boldsymbol{\Phi}^{-1}_l\mathbf{z}_l$ with $\boldsymbol{\Phi}_l=[\Phi_l(x^{[l]}_i,x^{[l]}_j)]_{i,j=1,\ldots,n_l}$ and $\mathbf{z}_l:=(f_l-f_{l-1})|_{\mathcal{X}_l}$. Then, a multi-level RKHS interpolator is introduced as follows, 
\begin{equation}\label{eq:emu}
    \hat{f}_L(x)=\sum^L_{l=1}\mathcal{P}_l(x).
\end{equation}

The resulting interpolator $\hat{f}_L(x)$ has some similarities to the co-kriging models \cite{kennedy2000predicting,le2014recursive,le2015cokriging}, which also assumes nested designs but models instead the function $( f_l - \rho_{l-1} f_{l-1} )$ via a GP prior using a Bayesian framework, where $\rho_{l-1}$ is an unknown parameter. 
A potential alternative model is the non-stationary GP proposed in \cite{tuo2014surrogate}.
One advantage of the interpolator \eqref{eq:emu},
as discussed in \cite{ehara2021}, is that it explicitly models for the bias between the exact simulation solution and its surrogate model, thus allowing for 
quantify the emulation accuracy in theory, which is developed in the next section. For further insight on the comparison between GP and RKHS interpolators, we refer readers to the works of \cite{wahba1990spline,lukic2001stochastic,scheuerer2013interpolation,kanagawa2018gaussian}, and for  potential extensions to GP emulators are discussed in Section \ref{sec:discussion}.

In the following, we make use of the Mat\'ern kernel $\Phi_l(x_i,x_j)=\phi_{\nu_l}(\|\Theta^{-1}_l(x_i-x_j)\|_2)$, where:
\begin{equation}\label{eq:maternkernel}
    \phi_{\nu_l}(r)=\frac{ 2^{1-\nu_l}}{\Gamma(\nu_l)}\left(r\sqrt{2\nu_l}\right)^{\nu_l}B_{\nu_l}\left(r\sqrt{2\nu_l}\right).
\end{equation}
Here, $\nu_l > 0$ is a smoothness parameter which controls differentiability of the interpolator, $\Theta_l$ is a diagonal $d\times d$ matrix of lengthscale parameters, $B_{\nu_l}$ is the modified Bessel function of the second kind, and $\Gamma$ is the gamma function. Mat\'ern kernels are widely used in the computer experiment \cite{santner2003design,gramacy2020surrogates} and spatial statistics \cite{stein2012interpolation} literature, and we will show later that such a kernel choice yields useful insights for characterizing the cost complexity performance of stacking designs. 

Of course, the  hyperparameters $\Theta_l$ and $\nu_l$ are unknown in practice and need to be estimated. These hyperparameters can be chosen to minimize the cross-validation error \cite{haaland2011accurate}. Although cross-validation methods are typically expensive to implement, the leave-one-out cross-validation (LOOCV) error of RKHS interpolators  can be expressed in a closed form \cite{haaland2011accurate}, which makes the computation more efficient. In particular, the LOOCV error of the interpolator $\mathcal{P}_l(x)$ is
\begin{equation}\label{eq:LOOCV}
    \frac{1}{n_l}\| \boldsymbol{\Lambda}_l^{-1}\boldsymbol{\Phi}_l^{-1}\mathbf{z}_l\|^2_2,
\end{equation}
where $\boldsymbol{\Lambda}_l$ is a diagonal matrix with the element $(\boldsymbol{\Lambda}_l)_{j,j}=(\boldsymbol{\Phi}_l^{-1})_{j,j}$, and the hyperparameters $\Theta_l$ and $\nu_l$ can be chosen by minimizing \eqref{eq:LOOCV}.

\section{Stacking designs}
\label{sec:nofidelity}
With the multi-level RKHS interpolator in hand, we now introduce the proposed stacking designs. We first define some notation. Let $\mathcal{N}_{\Phi}(\Omega)$ be the RKHS associated with a kernel $\Phi$, and let $\|g\|_{\mathcal{N}_{\Phi}(\Omega)}$ denote its RKHS norm for a function $g\in\mathcal{N}_{\Phi}(\Omega)$. In the following, we consider the refinement $(f_l-f_{l-1})$ to live on the RKHS $\mathcal{N}_{\Phi_l}(\Omega)$.

Suppose now, for each fidelity level $l$, there corresponds a measure of fidelity $\xi_l > 0$ quantifying how close the simulated response surface $f_l(\cdot)$ is to the \textit{exact solution}, which we denote as $f_\infty(\cdot)$. As $l\rightarrow\infty$, it is intuitive that $\xi_l\rightarrow 0$, meaning we approach the limiting solution as fidelity level increases. However, the exact solution $f_\infty(\cdot)$ typically cannot be computed numerically and must be approximated. In the case of finite element method (FEM), which is widely used for computer experiments, one such parameter for $\xi_l$ is the mesh size: a smaller mesh size $\xi_l$ results in higher mesh density and thus a more accurate simulator, at the cost of higher computation.


We now investigate the error in approximating the desired exact solution $f_\infty(x)$ with the multi-fidelity interpolator $\hat{f}_L(x)$ in \eqref{eq:emu}, which can be decomposed as 
\begin{equation}
    |f_{\infty}(x)-\hat{f}_L(x)|\leq\underbrace{|f_{\infty}(x)-f_L(x)|}_\text{simulation error}+\underbrace{|f_L(x)-\hat{f}_L(x)|}_\text{emulation error}.
    \label{eq:errdecomp}
\end{equation}
The first term corresponds to the \textit{simulation error}, which measures the discrepancy between the simulated solution $f_L(x)$ at fidelity level $L$ and the true/exact function $f_{\infty}(x)$. This error can be reduced by increasing $L$, the fidelity level $L$ of the simulator, or equivalently by reducing $\xi_l$. The second term represents \textit{emulation error} - the error for the multi-level interpolator given limited evaluations of the simulators.  This error can be reduced by increasing $n_L$, the sample size at the fidelity level $L$, or increase $n_{L-1}, \ldots, n_1$, the sample sizes at lower fidelity levels.

With this, the proposed stacking designs provide a batch sequential design scheme that aims to achieve the desired prediction accuracy of $\epsilon > 0$, i.e., $\|f_L-\hat{f}_L\|\leq \epsilon$. Here, $\|\cdot\|$ may be the $L_2$ or $L_\infty$ norm. This can be achieved by making both the simulation and emulation errors smaller than $\epsilon/2$, i.e.,
\begin{equation}
\|f_{\infty}-f_L\|\leq \epsilon/2\quad\text{and}\quad\|f_L-\hat{f}_L\|\leq\epsilon/2.
\label{eq:errdecomp2}
\end{equation}
We thus propose our stacking designs in two parts. In Section \ref{sec:samplesize}, we first present a sample size determination approach which bounds the emulation error $\|f_L-\hat{f}_L\|$ via closed-form expressions for sample sizes at each fidelity level. In Section \ref{sec:bias}, we then present a useful stopping rule on the maximum fidelity $L$ such that the simulation error bound on $\|f_{\infty}-f_L\|$ is satisfied. We then discuss a sequential algorithm for stacking designs in Section \ref{forwardAlgo}, and prove a novel complexity theorem that supports its performance in Section \ref{sec:complexitytheorem}.

\subsection{Emulation error control}\label{sec:samplesize}

Consider first the emulation error $\|f_L-\hat{f}_L\|$. We first present a useful bound on this error for the multi-level interpolator (proof in Supplementary Material \ref{append:theoremproofemulation}).

\begin{proposition}\label{thm:emulationaccuracy}
Suppose $\Omega$ is bounded and convex, $(f_l-f_{l-1}) \in \mathcal{N}_{\Phi_l}(\Omega)$. Assume that there exist constants $C_1,C_2,C_3,C_4,\nu_{\min}, \nu_{\max}>0$ such that $C_1\leq \|\Theta_l^{-1}\|_2 \leq C_2$, $C_3\leq \|\Theta_l\|_2 \leq C_4$, $\nu_{\min}\leq \nu_l\leq \nu_{\max}$, for all $l=1,...,L$. Then one can bound the prediction error of $\hat{f}_L(x)$ as:
\begin{equation}
|f_L(x)-\hat{f}_L(x)|\leq c_0\sum^L_{l=1}\|\Theta^{-1}_l\|^{\nu_l}_2h_{\mathcal{X}_l}^{\nu_l}\|f_l-f_{l-1}\|_{\mathcal{N}_{\Phi_l}(\Omega)}
\label{eq:errbd}
\end{equation}
for some constant $c_0>0$. Here, $h_{\mathcal{X}_l}$ is the fill distance \cite{wendland2004scattered} of the design $\mathcal{X}_l$, i.e., $h_{\mathcal{X}_l}=\sup_{x\in\Omega}\min_{x_u\in \mathcal{X}_l}\|x-x_u\|_2$.
\end{proposition}
This proposition nicely decomposes the prediction error of the multi-level interplator $\hat{f}_L$ into three distinct components at each fidelity level $l = 1, \ldots, L$. The first term, $\|f_l-f_{l-1}\|_{\mathcal{N}_{\Phi_l}(\Omega)}$, captures the size of the refinement with respect to its corresponding RKHS norm. This is quite intuitive, since a larger norm of the refinement $f_l - f_{l-1}$ is expected to induce greater error. The second term, $h_{\mathcal{X}_l}^{\nu_l}$, measures the quality of the design $\mathcal{X}_l$ in terms of how well it fills the design space $\Omega$. Note that a smaller fill distance $h_{\mathcal{X}_l}$ suggests that there are less ``gaps'' between design points \cite{mak2018minimax}, which in turn should reduce prediction error. The third term, $\|\Theta^{-1}_l\|^{\nu_l}_2$, captures the magnitude of the lengthscales $\Theta_l$. 
These three components provide the basis for the stacking sequential design method presented next.

We now wish to minimize the error bound in \eqref{eq:errbd} under a total budget on computational resources, to yield easy-to-evaluate expressions for determining sample sizes $n_l$ at each fidelity level $l = 1, \cdots, L$. 
Let $C_l$ be the computational cost (e.g., in CPU hours) for a single run of the simulator at fidelity level $l$. Note that, since higher-fidelity simulators are more computationally intensive, this implies that $0<C_1< C_2<\ldots< C_L$. 

From the experimental design perspective, an appealing design criterion for interpolation is \textit{quasi-uniformity} \cite{wendland2004scattered}, which ensures design points are uniformly placed over the design space $\Omega$. Specifically, denote $q_{\mathcal{X}} = \min_{i\neq j}\|x_i-x_j\|/2$, and a design $\mathcal{X}=\{x_i\}_{i=1}^n$ satisfying $h_{\mathcal{X}}/q_{\mathcal{X}}\leq c$, for some constant $c$, is called a quasi-uniform design. 
Such a design
satisfies the fill distance bound \cite{wendland2004scattered,muller2009komplexitat} $h_{\mathcal{X}} \leq c_1 n^{-1/d}$ for some constant $c_1 > 0$, where $n$ is the number of design points in $\mathcal{X}$. Quasi-uniformity has been widely studied in the literature \cite{fang1993number}, and there are a variety of designs which enjoy this property \cite{wynne2021convergence}. 
We thus restrict the designs $\mathcal{X}_l$ to be quasi-uniform for $l = 1, \cdots, L$.
The construction of such designs is discussed later in Section \ref{forwardAlgo}. Under this restriction, the error bound in \eqref{eq:errbd} reduces to
\begin{align}
|f_L(x)-\hat{f}_L(x)|&\leq c_0\sum^L_{l=1}c^{\nu_l}_1\|\Theta^{-1}_l\|^{\nu_l}_2n_l^{-\nu_l/d}\|f_l-f_{l-1}\|_{\mathcal{N}_{\Phi_l}(\Omega)}\nonumber\\
&\leq c_0c^*_1\sum^L_{l=1}\|\Theta^{-1}_l\|^{\nu_l}_2n_l^{-\nu_{\min}/d}\|f_l-f_{l-1}\|_{\mathcal{N}_{\Phi_l}(\Omega)},
\label{eq:errbd2}
\end{align}
where $c^*_1=\max_{l=1,\ldots,L}c_1^{\nu_l}$ and $\nu_{\min}=\min_{l=1,\ldots,L}\nu_l$.

Consider now the sample size determination problem, where we wish to minimize the error bound in \eqref{eq:errbd2} under the constraint of the total computational budget, $\sum_{l=1}^L n_l C_l$. This can be done by the method of Lagrange multipliers, which aims to find the saddle point of the Lagrangian function:
\begin{equation*}\label{eq:optimalsamplesize}
\sum^L_{l=1}\left(\|\Theta^{-1}_l\|^{\nu_l}_2n_l^{-\nu_{\min}/d}\|f_l-f_{l-1}\|_{\mathcal{N}_{\Phi_l}(\Omega)}+\lambda n_lC_l\right),
\end{equation*}
where $\lambda>0$ is the Lagrange multiplier. With some algebraic manipulations (by setting the gradient of the above function to zero), one can show that the optimal sample size for $n_l$ is:
\begin{equation}
   n_l= \mu r_l,\quad\text{ where }r_l=\left(\frac{\|\Theta^{-1}_l\|^{\nu_l}}{C_l}\|f_l-f_{l-1}\|_{\mathcal{N}_{\Phi_{l}}(\Omega)}\right)^{^{d/(\nu_{\min}+d)}},
   \label{eq:optss}
\end{equation} 
for some constant $\mu>0$. To ensure that $n_l$ is an integer, in our later implementation, we set it to the floor value of $\mu r_l$, i.e., $n_l=\lfloor\mu r_l\rfloor$.


The closed-form expression \eqref{eq:optss} reveals several useful insights for sample size determination in multi-fidelity experiments. First, with all things equal, we see that \eqref{eq:optss} allocates greater sample size $n_l$ for simulations with lower costs $C_l$ (i.e., lower fidelity simulations), which is intuitive. Second, note that \eqref{eq:optss} assigns greater sample size $n_l$ for fidelity levels where the refinement $f_l-f_{l-1}$ is more complex, whether that be in terms of a smaller lengthscale $\Theta_l$ or a larger RKHS norm. In particular, note that the RKHS norm of $f_l-f_{l-1}$ captures \textit{dissimilarities} of the simulators from fidelity level $l-1$ to level $l$. Thus, by minimizing \eqref{eq:optss}, our approach naturally factors in this dissimilarity information for optimal sample size allocation. 




There is still a free constant $\mu$, which we can set to achieve the desired emulation error $\|f_L-\hat{f}_L\|< \epsilon/2$. This parameter can be optimized as follows. By Theorem 11.14 of \cite{wendland2004scattered}, an alternative pointwise error bound of $\hat{f}_L(x)$ is
\begin{equation}\label{eq:powerfunctionbound}
    |f_L(x)-\hat{f}_L(x)|\leq\sum^L_{l=1}\sigma_l(x)\|f_l-f_{l-1}\|_{\mathcal{N}_{\Phi_l}(\Omega)},
\end{equation}
where $\sigma_l(x)$ is the so-called \textit{power function} of the form  
\begin{equation}\label{eq:powerfunction}
    \sigma^2_l(x)=\Phi_l(x,x)-\Phi_l(x,\mathcal{X}_l)\boldsymbol{\Phi}_l^{-1}\Phi_l(x,\mathcal{X}_l)^T,
\end{equation}
where $\Phi_l(x,\mathcal{X}_l)=\{\Phi_l(x,y)\}_{y\in \mathcal{X}_l}$. In contrast from the error bound in Theorem \ref{thm:emulationaccuracy}, the bound \eqref{eq:powerfunctionbound} does not depend on any constants, which allows for the following development. By the triangle inequality, $\|f_L-\hat{f}_L\|$ can be bounded by
\begin{equation}\label{eq:powerfunctionboundL2}
    \|f_L-\hat{f}_L\|\leq\sum^L_{l=1}\|\sigma_l\|\|f_l-f_{l-1}\|_{\mathcal{N}_{\Phi_l}(\Omega)}.
\end{equation}
To ensure the bound in \eqref{eq:powerfunctionboundL2} is less than the desired error tolerance $\epsilon / 2$, one can set the constant $\mu$ by solving the optimization problem
\begin{equation}\label{eq:giventolerance}
    \mu^*=\argmin_{\mu>0} \Bigg|\frac{\epsilon}{2}-\sum^L_{l=1}\|\sigma_l\|\|f_l-f_{l-1}\|_{\mathcal{N}_{\Phi_l}(\Omega)}\Bigg|\ \text{s.t.}\ \sum^L_{l=1}\|\sigma_l\|\|f_l-f_{l-1}\|_{\mathcal{N}_{\Phi_l}(\Omega)}\leq\frac{\epsilon}{2}.
\end{equation}
This optimization ensures that the error bound is equal (or close) to $\epsilon/2$, while the constraint ensures that it remains below $\epsilon/2$. Here, the dependency of the objective function on $\mu$ is via the term $\|\sigma_l\|$. This is because $\sigma_l(x)$ depends on the sample size $n_l$ and subsequently $\mu$ (recall $n_l=\lfloor\mu r_l\rfloor$), since the power function \eqref{eq:powerfunction} depends on the design points $\mathcal{X}_l$. To optimize for $\mu^*$ in \eqref{eq:giventolerance}, $\|\sigma_l\|$ can be approximated numerically via Monte Carlo integration \cite{caflisch1998monte} for $L_2$ norm and grid search optimization for $L_{\infty}$ norm. While one can alternatively set a fixed total cost budget (i.e., $\sum^L_{l=1}n_lC_l$) for determining $\mu$, we will focus here on achieving a desired predictive accuracy threshold $\epsilon$.

Finally, to optimize $\mu^*$, we also require knowledge of the RKHS norm $\|f_l-f_{l-1}\|_{\mathcal{N}_{\Phi_l}(\Omega)}$. Of course, the exact norm is unknown in implementation since it depends on $f_l$ and $f_{l-1}$. One can, however, approximate this term via its RKHS interpolator, which can be shown to equal to
$(\mathbf{z}_l^T\boldsymbol{\Phi}^{-1}_l\mathbf{z}_l)^{1/2}$ \cite{wendland2004scattered}. It should be noted that prior to optimization (i.e., before obtaining $\mathcal{X}_l$), we require information about the data $\mathbf{z}_l$ and the matrix $\boldsymbol{\Phi}_l$ in order to approximate the RKHS norm. These can be initially obtained via simulations on a pilot sample (we shall call this $\mathcal{X}_0$ later), then adaptively updated after collecting additional data from our stacking design (see Section \ref{forwardAlgo}). Similarly, the kernel parameters $\Theta_l$ and $\nu_l$, which are also required for the optimization \eqref{eq:giventolerance} and obtaining the sample size as in \eqref{eq:optss}, can also be adaptively estimated using the pilot sample $\mathcal{X}_0$  via cross-validation as mentioned in Section \ref{sec:model}. With these plug-in estimates within \eqref{eq:giventolerance}, the desired $\mu^*$ can then be efficiently obtained via simple dichotomous search \cite{ahlswede1987search}.


We note that the above sample size determination approach has several key distinctions from that in \cite{ehara2021}. First, our proposed sample sizes are determined by the desired prediction accuracy $\epsilon$ via \eqref{eq:giventolerance}, while theirs are controlled by a fixed total cost budget. This becomes important in the following subsection, where we leverage such sample sizes within a sequential stacking approach for controlling prediction error. Second, our approach makes use of an initial pilot sample to estimate the RKHS norm and other required parameters in \eqref{eq:optss}, which lends well to our later sequential design procedure. The approach in \cite{ehara2021}, by contrast, makes use of misspecified kernel functions, in particular, the rates proved in \cite{wang2020prediction}.

\subsection{Simulation error control via stacking}\label{sec:bias}

Consider next the simulation error $\|f_{\infty}-f_L\|$ in \eqref{eq:errdecomp2}, which concerns the numerical error of the simulator at fidelity level $L$. For many numerical simulators, this error can be bounded as
\begin{equation}\label{eq:numericbias}
   |f_{\infty}(x)-f_l(x)|\leq c_1(x)\xi^\alpha_l, \quad \text{for all} \; x \in \Omega,
\end{equation}
for some positive constants $\alpha$ and $c_1(x)$ depending on $x$. Recall that $\xi_l$ is a fidelity parameter which quantifies how close $f_l(\cdot)$ is to the exact solution $f_{\infty}(\cdot)$; the smaller $\xi_l$ is, the higher the fidelity of the simulator $f_l(\cdot)$. Equation \eqref{eq:numericbias} thus assumes that the simulation error is decaying polynomially in the fidelity parameter $\xi_l$. In the case of FEM with $\xi_l$ taken as the finite-element mesh size, it is well-known that the bound \eqref{eq:numericbias} holds under regularity conditions on the underlying solution (see, e.g.,  \cite{brenner2007fem,tuo2014surrogate}). Similarly polynomial decay rates have also been shown in a broad range of numerical simulators, e.g., for elliptical PDEs \cite{hundsdorfer2003numerical} and large-eddy simulations in fluid mechanics \cite{templeton2015calibration}.
We will thus assume the error bound in \eqref{eq:numericbias}, and leverage this for controlling simulation error via the stacking designs presented next.


Suppose the fidelity parameters $\{\xi_1, \xi_2, \cdots\}$ follow a geometric sequence for increasing fidelity levels, i.e., $\xi_l=\xi_0T^{-l},l\in\mathbb{N}^+$ for some integer $T\geq 2$. In the setting of FEM, where $\xi_l$ measures finite-element mesh size, $\xi_1, \xi_2, \cdots$ correspond to mesh sizes for increasing mesh refinements. The use of such a geometric sequence is motivated by multi-level Morte Carlo (MLMC) \cite{giles2008multilevel,giles2015multilevel}, which makes use of a similar sequence for time discretization of stochastic differential equations.

We now wish to ensure the simulation error satisfies the desired bound of $\|f_{\infty}-f_L\|\leq\epsilon/2$. Let us first assume a slightly stronger condition (compared to \eqref{eq:numericbias}):
\begin{align}\label{eq:highorderapprox}
f_{\infty}(x)-f_L(x)= c_1(x)\xi^{\alpha}_L+O(\xi_L^{\alpha+1}),
\end{align}
where $O(\xi_L^{\alpha+1})$ denotes a leading error term on the order of $\xi_L$ to the $(\alpha+1)$-th power. This assumption is common in the scientific computing literature, where it is referred to as the $\alpha$-order \textit{discretization error} to the mesh spacing parameter $\xi_L$ \cite{bect:hal-03096722,oberkampf2010verification}. Using the above sequence, $\xi_l=\xi_0T^{-l}$, and \eqref{eq:highorderapprox}, it follows that
\begin{align}\label{eq:highorderapprox2}
f_{\infty}(x)-f_{L-1}(x)=c_1(x)\xi^{\alpha}_LT^{\alpha}+O(\xi_L^{\alpha+1}).
\end{align}
By multiplying \eqref{eq:highorderapprox} by $T^{\alpha}$ and subtracting \eqref{eq:highorderapprox2}, we get
$$
f_{\infty}(x)=\frac{1}{T^{\alpha}-1}(T^{\alpha}f_L(x)-f_{L-1}(x))+O(\xi_L^{\alpha+1}).
$$
It thus follows that 
\begin{equation}\label{eq:biasuncertainty}
f_{\infty}(x)-f_L(x)=\frac{f_L(x)-f_{L-1}(x)}{T^{\alpha}-1},
\end{equation}
where terms of order $\xi^{\alpha+1}_L$ and higher are neglected.
Of course, the numerator $(f_L(x)-f_{L-1}(x))$ is unknown in implementation; we can, however, estimate it via its RKHS interpolator, namely, $\mathcal{P}_L(x)$. Combining everything together, the following criterion then serves as a check for whether the desired simulation error bound $\|f_{\infty}-f_{L}\|\leq\epsilon/2$ is met, 
\begin{equation}\label{eq:stoppingrule}
\frac{\|\mathcal{P}_L(x)\|}{T^{\alpha}-1}\leq\frac{\epsilon}{2}.    
\end{equation}
The above procedure extends a similar argument made in \cite{giles2008multilevel} for bounding approximation error in MLMC.


From \eqref{eq:stoppingrule}, the rate parameter $\alpha$ plays an important role for ensuring $\|f_{\infty}-f_{L}\|\leq\epsilon/2$. One way to set this parameter, as suggested in \cite{tuo2014surrogate}, is to infer $\alpha$ via the known error bound \eqref{eq:numericbias} from numerical analysis. For example, in FEM, if the interest lies in the integration of $f_\infty(\cdot)$ over a specific region, then $\alpha=2$ is suggested \cite{tuo2014surrogate}. If such prior information is not available from numerical analysis, then one could instead estimate this rate from data. In particular, using the error expansion \eqref{eq:highorderapprox}, we can write 
\begin{align*}
    f_l(x)&=f_{\infty}(x)+c_1(x)T^{(L-l)\alpha}\xi_L^{\alpha}+O(\xi_L^{\alpha+1}),\\
    f_{l-1}(x)&=f_{\infty}(x)+c_1(x)T^{(L-l+1)\alpha}\xi_L^{\alpha}+O(\xi_L^{\alpha+1}),\\
    f_{l-2}(x)&=f_{\infty}(x)+c_1(x)T^{(L-l+2)\alpha}\xi_L^{\alpha}+O(\xi_L^{\alpha+1}).
\end{align*}
Neglecting terms of $\xi_L^{\alpha+1}$ and higher, and subtracting $f_{l-1}$ from $f_l$ and $f_{l-2}$ from $f_{l-1}$ yields
\begin{align*}
    f_l(x)-f_{l-1}(x)&=c_1(x)T^{(L-l)\alpha}\xi_L^{\alpha}(1-T^{\alpha}),\quad\text{and}\\
    f_{l-1}(x)-f_{l-2}(x)&=c_1(x)T^{(L-l)\alpha}\xi_L^{\alpha}T^{\alpha}(1-T^{\alpha}),
\end{align*}
which gives 
$$
\alpha=\frac{\log\left(\frac{f_{l-1}(x)-f_{l-2}(x)}{f_{l}(x)-f_{l-1}(x)}\right)}{\log T} \quad\text{for}\quad l=3,\ldots,L.
$$
Thus, the parameter $\alpha$ can be estimated by the average value evaluated by the data: 
\begin{equation}\label{eq:alphaestimate}
\hat{\alpha}=\frac{1}{L-2}\sum^L_{l=3}\sum_{x\in\mathcal{X}_l}\frac{\log\left(\big|\frac{f_{l-1}(x)-f_{l-2}(x)}{f_{l}(x)-f_{l-1}(x)}\big|\right)}{n_l\log T},
\end{equation}
where the absolute value is used to ensure a positive value within the logarithm.

It is worth noting that the above developments have direct analogies in the scientific computing literature, in terms of the \textit{Grid Convergence Index} (GCI) method and the assessment of \textit{observed order of accuracy}. Such methods are commonly used to quantify discretization uncertainty in numerical models governed by partial differential equations. The foundation of these developments lies in a re-interpretation of Richardson's extrapolation procedure \cite{richardson1911ix}, as employed in the above derivations; further details can be found in Chapter 8 of \cite{oberkampf2010verification}. A comparative analysis between GCI and a Bayesian alternative for discretization uncertainty quantification \cite{tuo2014surrogate} is discussed in \cite{bect:hal-03096722}.

\subsection{Stacking design algorithm}\label{forwardAlgo}

We now combine the two error control approaches in Sections \ref{sec:samplesize} and \ref{sec:bias} into a sequential algorithm for stacking designs. From previous developments, there are two key properties that the design points in $\mathcal{X}_l$, $l = 1, \cdots, L$, should satisfy: (i) they should be nested over fidelity levels, i.e., $\mathcal{X}_L\subseteq\cdots\subseteq \mathcal{X}_1$, and (ii) for each fidelity level $l$, the design $\mathcal{X}_l$ should satisfy the quasi-uniformity property discussed in Section \ref{sec:samplesize}. One way to satisfy both properties is first choose a quasi-uniform sequence $\{z_i\}_{i=1}^\infty$ on the domain $\Omega$, then construct the multi-level designs as $\mathcal{X}_l = \{z_i\}_{i=1}^{n_l}$, $l = 1, \cdots, L$. 
Although there are methods for constructing quasi-uniform sequences (see, e.g., the low-dispersion sequences  \cite{yakowitz2000global,breger2018points}, such approaches typically have inflexible sample sizes or are not easily adaptable in a batch sequential manner (as needed for stacking designs). In our later implementation, we made use of Sobol' sequences \cite{sobol1967distribution}, which have been used empirically as sequential quasi-uniform designs (see, e.g., \cite{tuo2020kriging}). While it is unclear whether such designs achieve the optimal rates required for quasi-uniformity \cite{teckentrup2020convergence,wynne2021convergence}, it appears to yield good empirical performance in our numerical studies.

We now describe the proposed stacking design algorithm in detail. The algorithm begins by selecting an initial fidelity level $L=1$, then choosing an initial pilot design $\mathcal{X}_0$ of size $n_0$. Here, we suggest the initial sample size $n_0$ to be $5d \sim 10d$. We then iterate the following steps:
\begin{enumerate}
    \item Run the simulator at fidelity level $L$ and observe $f_L(x)$ at the pilot design $\mathcal{X}_0$. Estimate the  hyperparameters $\{\Theta_l\}_{l=1}^L$ and $\{\nu_l\}_{l=1}^L$ by minimizing the LOOCV error \eqref{eq:LOOCV}, and the RKHS norms $\{\|f_l-f_{l-1}\|_{\mathcal{N}_{\Phi_{l}}(\Omega)}\}_{l=1}^L$ using the approach in Section \ref{sec:samplesize}.
    \item Using these estimated parameters, compute the optimal sample sizes $n_l$ via \eqref{eq:optss} and \eqref{eq:giventolerance} for the current fidelity levels $l = 1, \cdots, L$. With this, construct designs $\mathcal{X}_l$ (with sample size $n_l$) to satisfy a nested structure, i.e., $\mathcal{X}_0 \subseteq \mathcal{X}_L\subseteq \mathcal{X}_{L-1}\subseteq\cdots\subseteq \mathcal{X}_1$. Run the simulators at these design points at their respective fidelity levels.
    \item If the number of fidelity levels $L \geq 3$, estimate the rate parameter $\alpha$ in \eqref{eq:numericbias} via regression (see Section \ref{sec:bias}). Using this estimate, test convergence via the stopping rule \eqref{eq:stoppingrule}.
    \item If convergence is not satisfied, iterate $L \leftarrow L + 1$ and repeat the above three steps. Otherwise, stop the batch sequential design and return the multi-level RKHS interpolator \eqref{eq:emu}.
\end{enumerate}
Note that the computed sample sizes from Step 2 aim to control the emulation error $|f_L(x)-\hat{f}_L(x)|$ (see Section \ref{sec:samplesize}), and the stopping rule in Step 3 aims to control the simulation error $|f_{\infty}(x)-f_L(x)|$ (see Section \ref{sec:bias}). The workflow of this batch sequential design algorithm is visualized in Figure \ref{fig:workflow}, and further details on the algorithm can be found in Supplementary Material \ref{alg:stackingdesign}.

By combining \eqref{eq:biasuncertainty} and \eqref{eq:powerfunctionbound} and replacing the refinement $(f_l-f_{l-1})$ with its interpolator $\mathcal{P}_l(x)$, an approximate pointwise error interval of $f_{\infty}(x)$ can be constructed as
\begin{equation}\label{eq:predictiveinterval}
   \hat{f}_L(x)\pm\left(\frac{|\mathcal{P}_L(x)|}{T^{\alpha}-1}+\sum^L_{l=1}\sigma_l(x)(\mathbf{z}_l^T\boldsymbol{\Phi}^{-1}_l\mathbf{z}_l)^{1/2}\right).
\end{equation}
These intervals can be used to quantify emulation uncertainty of the multi-level interpolator.

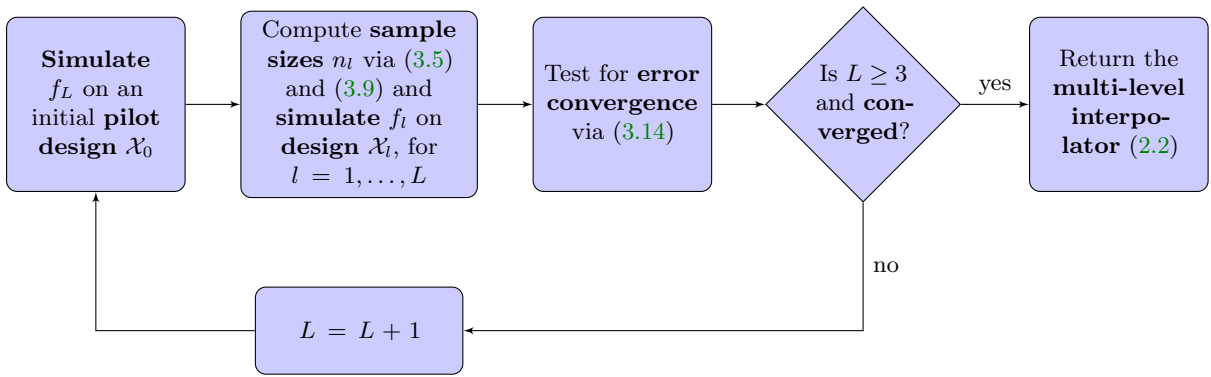
\begin{figure}[!t]   
\centering
\tikzstyle{decision} = [diamond, draw, fill=blue!20, 
    text width=4em, text badly centered, node distance=3.2cm, inner sep=0pt]
\tikzstyle{block} = [rectangle, draw, fill=blue!20, 
    text width=6.5em, text centered, rounded corners, minimum height=6em]
\tikzstyle{line} = [draw, -latex']
\tikzstyle{cloud} = [draw, ellipse,fill=red!20, node distance=3cm,
    minimum height=2em]
    
\begin{tikzpicture}[node distance = 3.2cm, auto]
\tikzstyle{every node}=[font=\footnotesize]
    \node [block, text width=5.5em] (init) {\textbf{Simulate} $f_L$ on an initial \textbf{pilot design} $\mathcal{X}_0$};
    \node [block, right of=init, text width=7.5em, node distance=3.5cm] (samplesize) {Compute \textbf{sample sizes} $n_l$ via \eqref{eq:optss} and \eqref{eq:giventolerance} and \textbf{simulate} $f_l$ on \textbf{design} $\mathcal{X}_l$, for $l=1,\ldots,L$}; 
    \node [block, right of=samplesize, text width=5.5em, node distance=3.5cm] (bias) {Test for \textbf{error convergence} via \eqref{eq:stoppingrule}};
    \node [decision, right of=bias] (decide) {Is $L \geq 3$ and \textbf{converged}?};
    \node [block, below of=samplesize, node distance=3cm, minimum width=2em, minimum height=3em] (update) {$L=L+1$};
    \node [block, right of=decide, node distance=3.4cm, text width=5.5em] (output) {Return the \textbf{multi-level interpolator} \eqref{eq:emu}};
    \path [line] (init) -- (samplesize);
    \path [line] (samplesize) -- (bias); 
    \path [line] (bias) -- (decide);  
    \path [line] (decide) |- node [near start] {no} (update);
    \path [line] (update) -| (init);
    \path [line] (decide) -- node {yes} (output);
\end{tikzpicture}
\caption{Visualizing the sequential workflow for the proposed stacking design algorithm. 
}
\label{fig:workflow}
\end{figure}


For the practical choice of the desired prediction accuracy $\epsilon$, one can begin with a large $\epsilon$ to conduct the  proposed stacking design, and if the prediction performance is not satisfactory (e.g., unsatisfactory predictions when comparing with validation simulations), the prediction accuracy can be subsequently improved by performing a ``post'' stacking design. More precisely, one can increase the precision by further selecting a smaller $\epsilon$ and iterating the above steps of the algorithm starting from the fidelity level $L$, at which the previous stacking design was terminated. This can be naturally done because the additional design points can be stacked on the previous stacking design.

\section{Cost complexity theorem}\label{sec:complexitytheorem}

With this in hand, we now investigate the computational cost for training data simulation to achieve the desired prediction error of $\epsilon$ in the following novel theorem. It should be noted that this theorem does not specify which design is used; it shows the \textit{existence} of multi-fidelity designs that achieve the asserted computational complexity with desired prediction error, and provides useful insights on conditions under which multi-fidelity emulation may be most effective. Such existence results are commonly encountered in the deterministic sampling literature (see, e.g., pg. 40 of \cite{dick2013high} and \cite{mak2018support}), and typically serve as a first step for design construction. Our theorem is also similar in spirit to the cost complexity theorems in \cite{giles2008multilevel} for multi-level Monte Carlo, extended to the multi-level interpolation setting at hand.

Let $C_l$ denote the computational cost required for a single simulation run at fidelity level $l$, and let $C_{\rm tot} = \sum_{l=1}^L n_l C_l$, which is the total computational budget for training data simulation. Our cost complexity theorem is stated as follows:
\begin{theorem}\label{thm:complexity}
Suppose $\Omega$ is bounded and convex, with $\Phi_l$ taken as the Mat\'ern kernel \eqref{eq:maternkernel}. Further suppose the smoothness parameters $\nu_l = \nu$ for each level $l$. 
Assume there exists positive constants $\alpha\geq \beta \nu/d,\beta,c_1,c_2,c_3$ and $c_4$ such that, for $l=1,\ldots,L$,
\begin{enumerate}
    \item the \textup{simulation error} $|f_{\infty}(x)-f_l(x)|$ is bounded as in \eqref{eq:numericbias}; furthermore, $c_1(x) \leq c_1$ for all $x\in \Omega$ and all $l=1,...,L$,
    \item the \textup{refinement function} $(f_l-f_{l-1}) \in\mathcal{N}_{\Phi_l}(\Omega)$ and there exists $v_l\in L_2(\Omega)$, such that $f_l(x)-f_{l-1}(x)=\int_{\Omega}\Phi_l(x,y)v_l(y){\rm{d}}y$, and $\bar{v}:=\sup_{l\in\mathbb{N}^+}\|v_l\|_{L_2(\Omega)}<+\infty$,
\item the \textup{kernel length-scale} parameters are bounded as $\|\Theta^{-1}_l\|_2<c_2$,
\item the \textup{designs} $\mathcal{X}_l$ are quasi-uniform, i.e., $h_{\mathcal{X}_l}<c_3n_l^{-1/d}$,
\item the \textup{computational cost} $C_l$ is bounded as $C_l\leq c_4\xi_l^{-\beta}$.
\end{enumerate}
Assuming an error tolerance of $0 < \epsilon<e^{-1}$, there then exist choices of $L$ and $n_1, \cdots, n_L$ for which the multi-level interpolator achieves the desired prediction bound
$$
|f_{\infty}(x)-\hat{f}_L(x)|\leq \epsilon, \quad x \in \Omega,
$$
with a \textup{total computational cost} $C_{\rm tot}$ bounded by
\begin{equation}
\begin{aligned}
C_{\rm tot} &\leq \left\{\begin{array}{ll} c_5 \epsilon^{-\frac{d}{\nu}},  &  \frac{\alpha}{\beta}>\frac{2 \nu}{d},\\
c_5\epsilon^{-\frac{d}{\nu}}|\log \epsilon |^{1+\frac{d}{\nu}}, &  \frac{\alpha}{\beta}=\frac{2 \nu}{d},\\
c_5\epsilon^{-\frac{d}{\nu}-\frac{2\beta \nu-\alpha d}{2\alpha(\nu+d)}}, &  \frac{\alpha}{\beta}<\frac{2 \nu}{d}.\end{array}\right.
\end{aligned}
\label{eq:compcost}
\end{equation}
where $c_5$ is a positive constant.
\end{theorem}

Note that Condition 2 implies that the extended function $(f_l-f_{l-1})_e\in H^{2\nu+d}(\mathbb{R}^d)$. This higher-order smoothness requirement allows for improved convergence rates in our specific context; see \cite{tuo2020improved} for further details. The proof of Theorem \ref{thm:complexity} is given in Supplementary Material  \ref{append:theoremproof}. The underlying principle of the proof is to require $L$ to be
\begin{equation*}
L=\bigg\lceil{\frac{\log(2c_1\xi_0^{\alpha}\epsilon^{-1})}{\alpha\log T}}\bigg\rceil
\end{equation*}
to ensure $|f_{\infty}(x)-f_L(x)|\leq\epsilon/2$, where $\lceil \cdot\rceil$ rounds up to the nearest integer and $c_1=\sup_{x\in\Omega}c_1(x)$,  and then select the optimal $n_1,\ldots,n_L$:
\label{eq:selectLandnl}
\begin{equation}
\quad n_l \propto \xi_l^{\frac{(\alpha+2\beta)d}{2(\nu+d)}},
\end{equation}
and a constant of proportionality is chosen so that $|f_L(x)-\hat{f}_L(x)|\leq\epsilon/2$. In practice, the value of $L$ is unknown due to the presence of the unknown constant $c_1$. This is where our proposed stacking design comes into play in practice, as it can be utilized to determine the value of $L$ effectively.

While this theorem is quite involved, it provides several novel and useful insights on the multi-fidelity design problem. By \eqref{eq:selectLandnl},  for a given fidelity level $l$,  
it follows that the computational work $n_l\xi_l^{-\beta}$ satisfies
\begin{equation}
    \quad n_l\xi_l^{-\beta} \propto \xi_l^\frac{\alpha d-2\beta \nu}{2(\nu+d)}, \quad l = 1, \cdots, L,
    \label{eq:propto}
\end{equation}
where $\xi_l$ is again the fidelity parameter for level $l$. From this, a key factor for determining how much of the total budget $C_{\rm tot}$ to allocate to each fidelity is whether the numerator of the last term $\alpha d - 2 \beta \nu > 0$, or equivalently, the factor $\alpha/\beta > 2\nu/d$. When $\alpha/\beta > 2\nu/d$, one can see from \eqref{eq:propto} that much of the budget $C_{\rm tot}$ is expended on the levels with lower fidelities, i.e., with coarser mesh densities. Conversely, when $\alpha/\beta < 2\nu/d$, much of the budget will be allocated to levels with higher fidelities, i.e., with denser mesh densities. 

One can further glean intuition on the terms $\alpha/\beta$ and $2\nu/d$ in this comparison. Recall that the parameters $\alpha$, $\beta$, $\nu$ and $d$   correspond to the rate parameter for simulator error convergence (see \eqref{eq:numericbias}), the rate of increase in computational cost $C_l$ as fidelity increases,  the smoothness of the refinement function $(f_l-f_{l-1})$, and the number of input parameters, respectively. One can thus interpret the first fraction ${\alpha}/{\beta}$ as the rate of simulator error reduction over the rate of computational cost increase as fidelity increases. Similarly, the second fraction ${2\nu}/{d}$ can be interpreted as the rate of convergence for the RKHS interpolator (see, e.g., \cite{wendland2004scattered}). Thus, when ${\alpha}/{\beta}$ exceeds this rate of convergence for the RKHS interpolator (due to a combination of (i) and (ii)), the design procedure would shift more computational resources towards lower fidelity simulation runs, which is quite intuitive.

To further explore this idea, we compare next the cost complexity rate in Theorem \ref{thm:complexity} with the corresponding rate if our emulator were trained using only high-fidelity simulation data. Note that the fidelity level chosen for this latter high-fidelity emulator may be different from the high-fidelity level $L$ for the multi-fidelity interpolator; to distinguish this, we will use fidelity level $H$ with corresponding fidelity parameter $\xi_H$. For this high-fidelity emulator, its predictor is given by the RKHS interpolator of $f_H\rvert_{\mathcal{X}_H}$, i.e.,
\begin{align}\label{eq:emulatorsingle}
\hat{f}_H(x)=\Phi_H(x,\mathcal{X}_H)\boldsymbol{\Phi}^{-1}_H\mathbf{y}_H,
\end{align}
where $\mathbf{y}_H:=f_H|_{\mathcal{X}_H}$ is the response vector simulated at fidelity level $H$. The following corollary shows a similar cost complexity result for this high-fidelity RKHS interpolator:

\begin{corollary}\label{cor:singlefidelity}
Assume that
\begin{enumerate}
    \item there exist some $\xi_H > 0$ and $0<\epsilon<1$ for which $\left(\epsilon/2\right)^{1+\frac{\alpha d}{2\nu\beta}}\leq c_1\xi_H^\alpha\leq\epsilon/2$, where $c_1=\sup_{x\in\Omega}c_1(x)$,
    \item the high-fidelity \textup{response surface} $f_H \in\mathcal{N}_{\Phi_H}(\Omega)$ and the extended function $(f_H)_e\in H^{2\nu+d}(\mathbb{R}^d)$,
    \item there exists a positive constant $c_2$ such that the \textup{kernel length-scale} parameter is bounded as $\|\Theta^{-1}_H\|_2<c_2$.
\end{enumerate}
There then exists a sample size $n_H$ for which the high-fidelity emulator \eqref{eq:emulatorsingle} achieves the desired prediction bound
$$
|f_{\infty}(x)-\hat{f}_H(x)|\leq \epsilon, \quad x \in \Omega,
$$
with a \textup{total computational cost} $C_H$ bounded by 
\begin{equation}
C_H\leq c_h\epsilon^{-\frac{\beta}{\alpha}-\frac{d}{2\nu}},
\label{eq:highfid}
\end{equation}
where $c_h$ is a positive constant.
\end{corollary}

By comparing the above rate with the complexity rate \eqref{eq:compcost} for the multi-level interpolator (Theorem \ref{thm:complexity}), one can gain illuminating insights on when multi-fidelity emulation improves upon standard high-fidelity RKHS interpolators. When $\alpha / \beta < 2 \nu / d$ (the same ratio compared earlier), the multi-level interpolator rate improves upon the high-fidelity interpolator rate, and conversely when $\alpha / \beta \geq 2 \nu / d$, the high-fidelity rate is quicker than the multi-level rate. Such a condition is intuitive and can be reasoned from the rate parameters $\alpha$ and $\beta$. For $\alpha$, take the limiting setting of $\alpha \rightarrow \infty$, such that the simulation error \eqref{eq:numericbias} decreases \textit{rapidly} to zero as fidelity increases. In this setting, it is intuitive that a high-fidelity interpolator (which relies solely on such high-accuracy runs) would outperform the multi-level interpolator; this is then affirmed by the fact that the condition $\alpha / \beta \geq 2 \nu / d$ is satisfied. For $\beta$, take the limiting setting of $\beta \rightarrow 0$, such that the computational cost $C_l$ grows \textit{slowly} as fidelity increases. In this case, it makes sense that a high-fidelity interpolator (which would not be costly) would outperform the multi-level interpolator given a fixed budget; this is again affirmed by the condition $\alpha / \beta \geq 2 \nu / d$. Similar conclusions also hold in reverse: when $\alpha \rightarrow 0$ or $\beta \rightarrow \infty$, i.e., when the simulation error \eqref{eq:numericbias} decreases \textit{rapidly} to zero or the cost $C_l$ grows \textit{rapidly} as fidelity increases, analogous reasoning can be used to explain why the multi-level interpolator would be more preferable than the high-fidelity interpolator given a fixed cost budget. In this view, Theorem \ref{thm:complexity} provides a novel perspective on when multi-fidelity modeling improves upon high-fidelity modeling for emulation. We note that the analysis is based on the assumption of a small error $\epsilon$ that satisfies  $\epsilon<2c_1\xi_1^{\alpha}$, which ensures the necessity of at least two fidelity levels for the multi-fidelity computer experiments in Theorem \ref{thm:complexity}.

It should be noted that, in the above analysis, it is assumed that one knows the high-fidelity fidelity parameter $\xi_H$ such that Condition 1 in Corollary \ref{cor:singlefidelity} is satisfied. For practical implementation, such a fidelity parameter is typically not known, and a misspecification of this $\xi_H$ can lead to a worse cost complexity rate than what is guaranteed by Corollary \ref{cor:singlefidelity} for the high-fidelity interpolator. The proposed stacking design gets around this issue of ``fidelity misspecification'', by employing a sequential sampling approach for determining the number of levels $L$ and corresponding sample sizes $n_1, \ldots, n_L$ to achieve the desired error tolerance.

Finally, we note that the above analysis is based on the established \textit{upper} bound on computational cost $C_{\rm tot}$ in \eqref{eq:compcost}. This might be made more concrete with a matching \textit{lower} bound on $C_{\rm tot}$; this would preclude the existence of designs with same accuracy but lower cost complexity, and provide a tight cost complexity bound for analysis. However, such a lower bound does not appear to be straight-forward, 
and we thus defer this as future work. The employed upper-bound cost analysis has been widely used in the analysis of multi-level Monte Carlo methods \cite{giles2008multilevel,giles2015multilevel}, where useful insights have been gleaned; our analysis above to do the same for the related problem of multi-level emulation.

\section{Numerical Experiments}\label{sec:numericstudy}

We now investigate a suite of numerical experiments to examine the proposed stacking designs, in particular, its predictive performance and its ability to achieve a desired error tolerance. Section \ref{sec:synthetic} explores a synthetic example, Section \ref{sec:casestudy} investigates an application to Poisson's equation, and Section \ref{sec:casestudy2} considers an application for thermal stress analysis of a jet engine turbine blade. The latter two problems involve partial differential equation (PDE) systems which are numerically solved via finite element modeling. These experiments are all initialized with an initial design $\mathcal{X}_0$ of size $n_0=5d$. All experiments are performed on a MacBook Pro laptop with Apple M1 Max Chip and 32Gb of RAM.





\subsection{Multi-fidelity Currin function}\label{sec:synthetic}
\begin{figure}[!t]
    \centering
    \includegraphics[width=\textwidth]{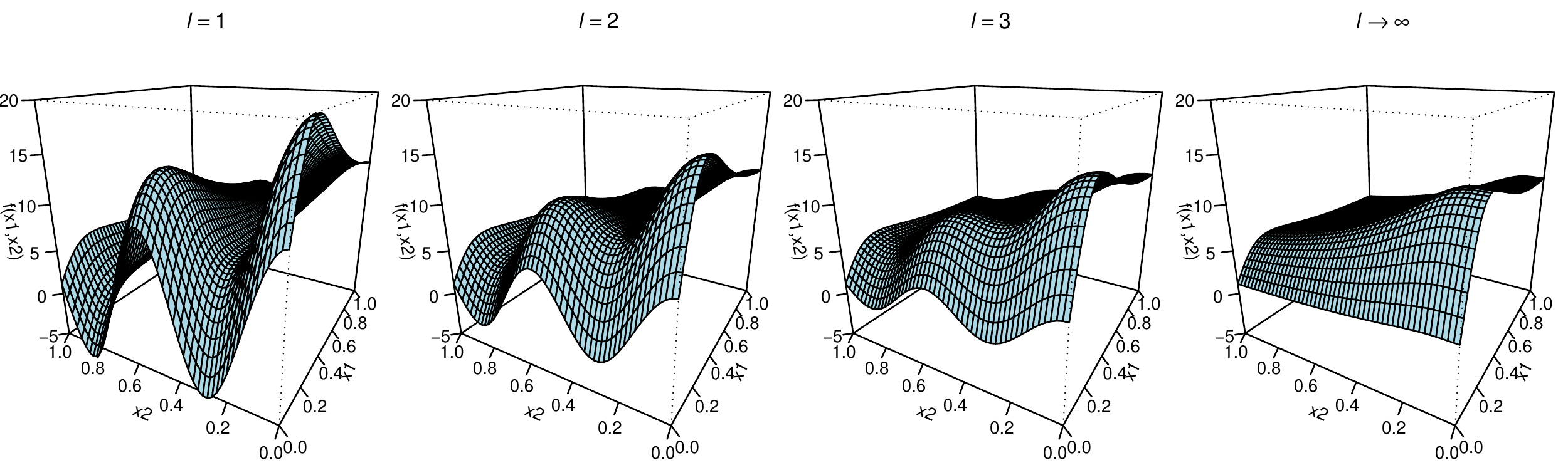}
    \caption{Visualizing the multi-fidelity Currin function $f_l(x_1,x_2)$ for $l=1,2,3$ and the limiting (highest-fidelity) Currin function $f_{\infty}(x_1,x_2)$.}
    \label{fig:synthetic_illustration}
\end{figure}

We first consider the following two-dimensional multi-fidelity Currin function:
\begin{equation}
f_l(x_1,x_2)=f_{\infty}(x_1,x_2)+\xi_l^{\alpha}\exp(-1.4x_1)\cos(3.5\pi x_2).
\label{eq:synth}
\end{equation}
Here, $f_{\infty}(x_1,x_2)$ is the Currin test function \cite{currin1988bayesian} which we take to be our limiting highest-fidelity setting:
$$
f_{\infty}(x_1,x_2)=\left[1-\exp\left(-\frac{1}{2x_2}\right)\right]\frac{2300x_1^3 + 1900x_1^2 + 2092x_1 + 60}{100x_1^3 + 500x_1^2 + 4x_1 + 20}, \quad (x_1,x_2)\in[0,1]^2=\Omega.
$$
The remaining term in \eqref{eq:synth} is the discrepancy term, which converges to zero as fidelity parameter $\xi_l$ increases and makes \eqref{eq:synth} satisfy the inequality of \eqref{eq:numericbias}. 
In the following, we set $\alpha=1$ and $\xi_l=\xi_0T^{-l}=16\times 2^{-l}$, and set the computational cost to be $C_l=4^l$. We assume that $\alpha$ is unknown which needs to be estimated. Figure \ref{fig:synthetic_illustration} shows the synthetic functions $f_l(x_1,x_2)$ for fidelity levels $l=1,2,3$ and the limiting function $f_{\infty}(x_1,x_2)$. With this, we then applied the stacking design algorithm from Section \ref{forwardAlgo}, with the desired prediction accuracy is set as $\epsilon=1$ in $L_2$ norm. 

\begin{figure}[!t]
    \centering
    \includegraphics[width=\textwidth]{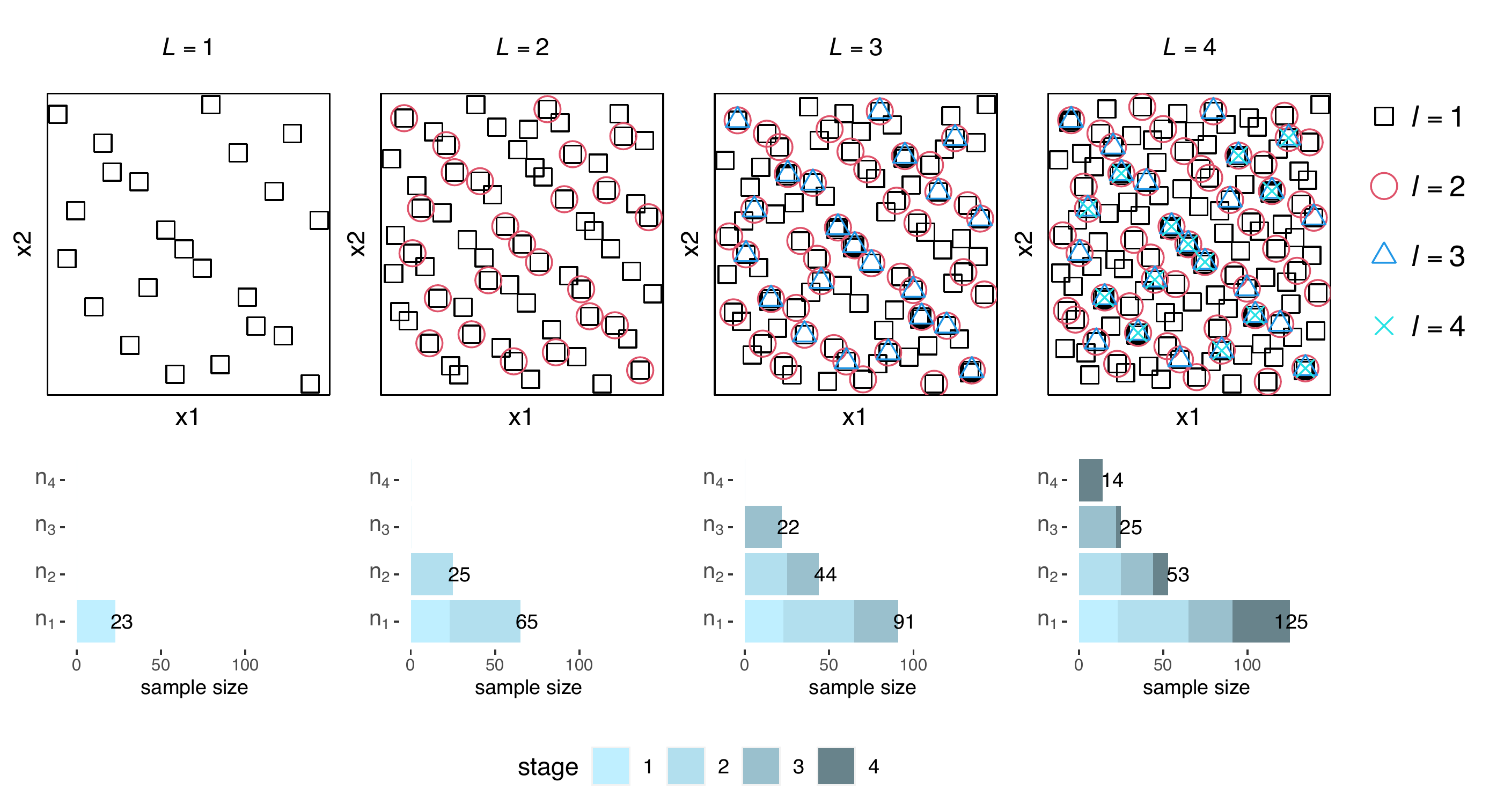}
    \caption{(Top) The proposed stacking designs at $L=4$ fidelity levels, over four batch sequential design stages (from left to right). (Bottom) The corresponding sample sizes at the $L=4$ fidelity levels, over four batch sequential design stages (from left to right).}
    \label{fig:synthetic_design}
\end{figure}

The stacking design begins with $L=1$, which requires $n_1=23$ design points on the lowest fidelity simulator $f_1$ (left panel of Figure \ref{fig:synthetic_design}) to ensure that the estimated emulation error bound \eqref{eq:powerfunctionboundL2} of $\|f_1-\hat{f}_1\|_{L_2(\Omega)}$ is smaller than $\epsilon/2 = 0.5$; this is summarized in the $L=1$ column in Table \ref{tab:synthetic2d}. In the next step with $L=2$, we then add on an additional 42 design points for the lower-fidelity simulator $f_1$, yielding a total of $n_1=65$ runs on $f_1$. We then add $n_2=25$ design points on a higher-fidelity simulator $f_2$ (i.e., fidelity level $l=2$), which is then ``stacked'' on top of the lower-fidelity design (second panel from the left of Figure \ref{fig:synthetic_design}). With the designs conducted at these two fidelity levels, the estimated emulation error bound $\|f_2-\hat{f}_2\|_{L_2(\Omega)}$ comes to 0.455, which is again less than $\epsilon / 2 = 0.5$. We then repeat this iteratively for increasing fidelity levels $L=3$ and $L=4$, after which the stopping rule \eqref{eq:stoppingrule} is satisfied and the procedure is terminated. Figure \ref{fig:synthetic_design} visualizes the experimental designs and corresponding samples sizes at each step.


To evaluate the simulation error bound in the stopping rule \eqref{eq:stoppingrule}, the rate parameter $\alpha$ needs to be estimated from data; this can be done via \eqref{eq:alphaestimate} when $L\geq 3$. The estimates of $\alpha$ (reported in Table \ref{tab:synthetic2d}) are precisely equal to the true parameter $\alpha=1$. With the estimate of $\alpha$ at $L=3$, the simulation error bound \eqref{eq:stoppingrule} gives 0.798, which is greater than $\epsilon/2$ and thus the batch sequential design continues. With the estimate of $\alpha$ at $L=4$, the simulation error estimate is less than $\epsilon/2 = 0.5$, thus the stopping rule is satisfied and the design procedure stops. Table \ref{tab:synthetic2d} shows the estimated upper bounds for simulation and emulation errors, both of which need to be smaller than $\epsilon/2=0.5$ for the procedure to stop. The $L_2$-error of the final multi-level interpolator (estimated via Monte Carlo integration) is $\|f_{\infty}-\hat{f}_4\|_{L_2(\Omega)}= 0.53$, which is indeed smaller than the desired prediction accuracy of $\epsilon=1$. This shows that the proposed stacking designs, by increasing fidelity levels and stacking design points in a sequential fashion, can indeed satisfy the desired error bound.

\begin{table}[!t]
    \centering
    \begin{tabular}{ccccccc}
        \toprule
         & $L=1$  &  $L=2$ & $L=3$  & $L=4$ \\
        \midrule
        Fidelity parameter & $\xi_1=8$ &$\xi_2=4$ & $\xi_3=2$ & $\xi_4=1$ \\
        Cost per run & $C_1=4$ &$C_2=16$ & $C_3=64$ & $C_4=256$ \\
        \midrule
        Bound of $\|f_{\infty}-f_{L}\|_{L_2(\Omega)}$  & NA & NA & 0.798 &  \textbf{0.448}\\
         &&&($\hat{\alpha}=1$)& ($\hat{\alpha}=1$)\\
      Bound of $\|f_L-\hat{f}_{L}\|_{L_2(\Omega)}$  & \textbf{0.483} & \textbf{0.455} &  \textbf{0.324} &  \textbf{0.474}\\
        \bottomrule
    \end{tabular}
    \caption{The estimated simulation and emulation error bounds (see \eqref{eq:stoppingrule} and \eqref{eq:powerfunctionboundL2}, respectively) at each design stage for the multi-fidelity Currin experiment, with estimated rate parameter $\hat{\alpha}$ at stages $L=3$ and $L=4$. Bolded numbers indicate the error is less than $\epsilon/2$, where $\epsilon=1$ is the desired error tolerance. }
    \label{tab:synthetic2d}
\end{table}

To benchmark against the state-of-the-art, we further implemented the sequential design strategy in \cite{le2015cokriging} using the above set-up, with code provided in the Supplementary Materials of their paper. This comparison serves \textit{solely} as a benchmark for gauging predictive accuracy of our approach; the above existing sequential design does not target a desired predictive accuracy $\epsilon > 0$, which is the key focus of our method. As the provided code was specifically designed for two fidelity levels (and is difficult to generalize for more levels), we uniformly sampled two values of $\xi_l$ (from 1 to 8) as the two fidelity levels for simulations, then performed the sequential design using the same total computational cost as our method (which is 6532) as a stopping criterion. This experiment was replicated 100 times. The average emulation $L_2$-error using the above existing sequential designs is 1.59 (with standard deviation of 0.73), which is higher than the error using the proposed stacking designs (0.53); thus, our approach appears to offer better (or at least comparable) predictions to this state-of-the-art approach.

\begin{figure}[h]
    \centering
    \includegraphics[width=\textwidth]{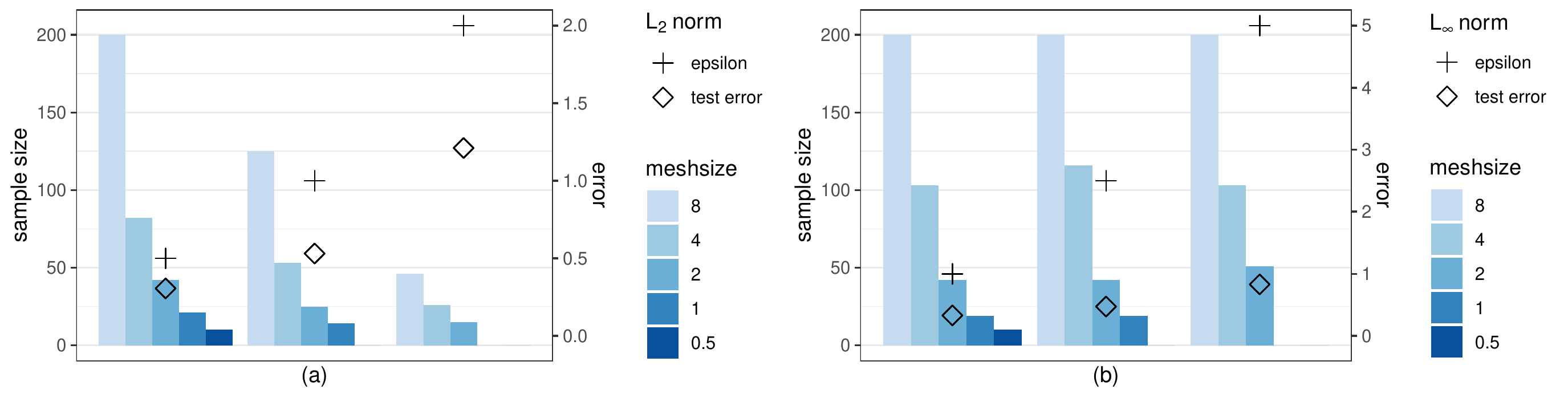}
    \caption{Visualizing the allocated sample sizes from stacking designs and the corresponding $L_2$ (left) and $L_\infty$ (right) test errors (marked $\diamondsuit$) at various error tolerances (marked $+$) for the multi-fidelity Currin experiment. }
    \label{fig:synthetic_epsilon}
\end{figure}

Finally, we explore the performance of stacking designs for different choices of error tolerance $\epsilon$, using both $L_2$ and $L_\infty$ norms. Figure \ref{fig:synthetic_epsilon} visualizes the sample sizes (at each fidelity level) and the corresponding errors of the final multi-level interpolator. We see that, for different $\epsilon$ and different norms, the proposed stacking designs can consistently yield prediction errors which are smaller than the desired error tolerance $\epsilon$, which is as desired.

\subsection{Poisson's equation}\label{sec:casestudy}
\begin{figure}[!t]
    \centering
    \includegraphics[width=\textwidth]{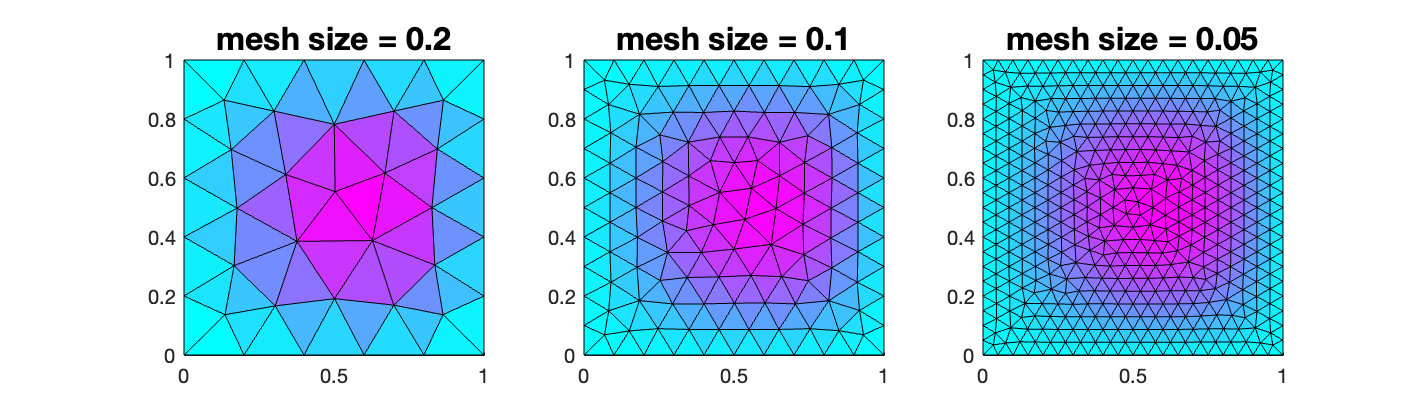}
    \caption{Visualizing the FEM solutions of Poisson's equation \eqref{eq:poisson} with three different mesh sizes.}
    \label{fig:mesh_demo}
\end{figure}

Next, we explore the performance of stacking designs for emulating an elliptical PDE system. The system of interest is modeled using Poisson's equation on a square membrane \cite{evans2010partial}, which has broad applicability in electrostatics and fluid mechanics. This can be represented by the PDE:
\begin{equation}
    \Delta u=(x^2-2\pi^2)e^{x z_1}\sin(\pi z_1)\sin(\pi z_2)+2x\pi e^{x z_1}\cos(\pi z_1)\sin(\pi z_2),\quad (z_1,z_2)\in D,
    \label{eq:poisson}
\end{equation}
where $u(z_1,z_2)$ is the solution of interest, $\Delta={\partial^2}/{\partial z_1^2}+{\partial^2}/{\partial z_2^2}$ is the Laplace operator, $D\in[0,1]\times[0,1]$, and $x\in\Omega=[-1,1]$. One then imposes the Dirichlet boundary condition $u=0$ on the boundary $\partial D$. Following \cite{tuo2014surrogate}, FEM is used to solve this system numerically. In our implementation, we make use of the Partial Differential Equation Toolbox of \cite{MATLAB:R2021b} to create the geometry and mesh. In the toolbox, one can specify the mesh size for the numerical solver; Figure \ref{fig:mesh_demo} visualizes the numerical solutions of \eqref{eq:poisson} with three different choices of mesh sizes.



Here, the response of interest is taken as the integral of the solution  $f_{\infty}(x)=\int u(z_1,z_2) {\rm{d}}z_1{\rm{d}}z_2$. It can be shown that this solution has the analytical form \cite{tuo2014surrogate}:
\begin{equation*}
    f_{\infty}(x)=\frac{2(e^x+1)}{x^2+\pi^2}.
\end{equation*}
Of course, in practical problems, one does not typically have such closed-form solutions; this framework was chosen to allow for easy validation of the fitted emulator model. For the stacking designs, we again define a geometric sequence of fidelity parameters (here, mesh sizes) $\xi_l=\xi_0T^{-l}=0.4\times 2^{-l}$, $ l=1,2,\ldots$. We then set a desired prediction accuracy of $\epsilon=0.05$ for the $L_{\infty}$-norm. 

\begin{figure}[!t]
    \centering
    \includegraphics[width=\textwidth]{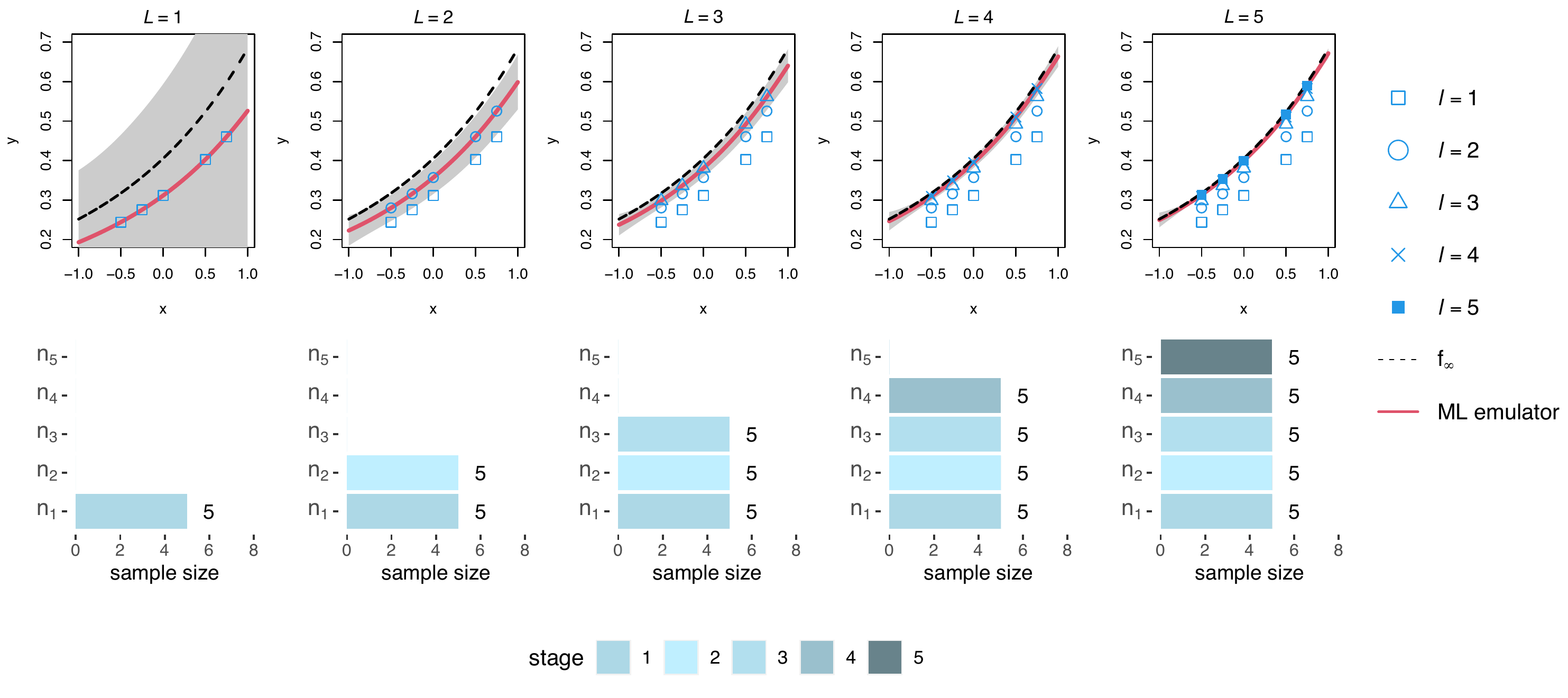}
    \caption{Visualizing the stacking design procedure for the Poisson's equation experiment. (Top) The stacking design points (dots) and the fitted multi-level interpolator with pointwise error intervals obtained via \eqref{eq:predictiveinterval} (gray shaded region) over five batch sequential design stages (from left to right). (Bottom) The corresponding sample sizes at the $L=5$ fidelity levels, over five batch sequential design stages (from left to right). }
    \label{fig:nofidelty_5d_proportion}
\end{figure}

Figure \ref{fig:nofidelty_5d_proportion} shows the fitted multi-level interpolator  and pointwise error intervals (upper panels) obtained via \eqref{eq:predictiveinterval}, with its corresponding sample sizes at each fidelity level (bottom panels) at each step of the stacking design process. Table  \ref{tab:possion_biases} shows the estimated emulation and simulation error bounds. 
In the first step with $L=1$, the lowest fidelity simulation (with mesh size $0.2$) is run with a design of size $n_1=5$. From the left panels in Figure \ref{fig:nofidelty_5d_proportion}, the resulting fitted multi-level interpolator $\hat{f}_1$ appears to be biased with very wide error bounds. 
In the third step ($L=3$), additional design points are stacked on fidelity levels 2 and 3 ($n_1=5,n_2=5,n_3=5$), and the simulation error rate parameter $\alpha$ is then estimated via \eqref{eq:alphaestimate}. With this estimated parameter, the estimated simulation error bound \eqref{eq:stoppingrule} evaluates to $0.047$ (see Table  \ref{tab:possion_biases}), which is being greater than $\epsilon/2 = 0.025$. The process thus continues until the fifth step ($L=5$), in which both the simulation and emulation error bounds become smaller than $\epsilon/2$. From the right panels in Figure \ref{fig:nofidelty_5d_proportion}, the resulting fitted emulator $\hat{f}_5$ appears to yield an accurate prediction with narrow error bounds. The $L_{\infty}$-error of this final multi-level interpolator (estimated via grid search optimization) is $\|f_{\infty}-\hat{f}_5\|_{L_\infty(\Omega)}=0.013$, which is smaller than the desired prediction accuracy $\epsilon=0.05$. This again shows that the proposed stacking designs, by iteratively increasing fidelity levels and stacking design runs, can yield the desired error tolerance. We again compare with the sequential design in \cite{le2015cokriging}. The experimental setup is similar to that described in Section \ref{sec:synthetic}. The results indicate that, at the same computational cost, the $L_\infty$-error of the emulator using the sequential designs is 0.058 (with standard deviation of 0.035), which is higher than ours (which is 0.013).



\begin{table}[h!]
    \centering
    \begin{tabular}{cccccccc}
        \toprule
         & $L=1$  &  $L=2$ & $L=3$  & $L=4$ & $L=5$\\
        \midrule
        Mesh size & $\xi_1=0.2$ &$\xi_2=0.1$ & $\xi_3=0.05$ & $\xi_4=0.025$ & $\xi_5=0.0125$\\
        Cost per run (sec.) & $C_1=0.18$ &$C_2=0.19$ & $C_3=0.23$ & $C_4=0.27$ &  $C_5=0.55$\\
        \midrule
        Bound of $\|f_{\infty}-f_{L}\|_{L_{\infty}(\Omega)}$ & NA & NA & 0.047 & 0.029 & \textbf{0.007}  \\
         &&&($\hat{\alpha}=0.92$)& ($\hat{\alpha}=0.84$)& ($\hat{\alpha}=1.07$)\\
       Bound of $\|f_L-\hat{f}_{L}\|_{L_{\infty}(\Omega)}$  & \textbf{0.008} & \textbf{0.011} & \textbf{0.014} & \textbf{0.015} & \textbf{0.015}\\
        \bottomrule
    \end{tabular}
    \caption{The estimated simulation and emulation error bounds (see \eqref{eq:stoppingrule} and \eqref{eq:powerfunctionboundL2}, respectively) at each design stage for the Poisson's equation experiment, with estimated rate parameter $\hat{\alpha}$ at stages $L=3,4$ and $5$. Bolded numbers indicate the error is less than $\epsilon/2$, where $\epsilon=0.05$ is the desired error tolerance.}
    \label{tab:possion_biases}
\end{table}

\begin{figure}[h]
    \centering
    \includegraphics[width=\textwidth]{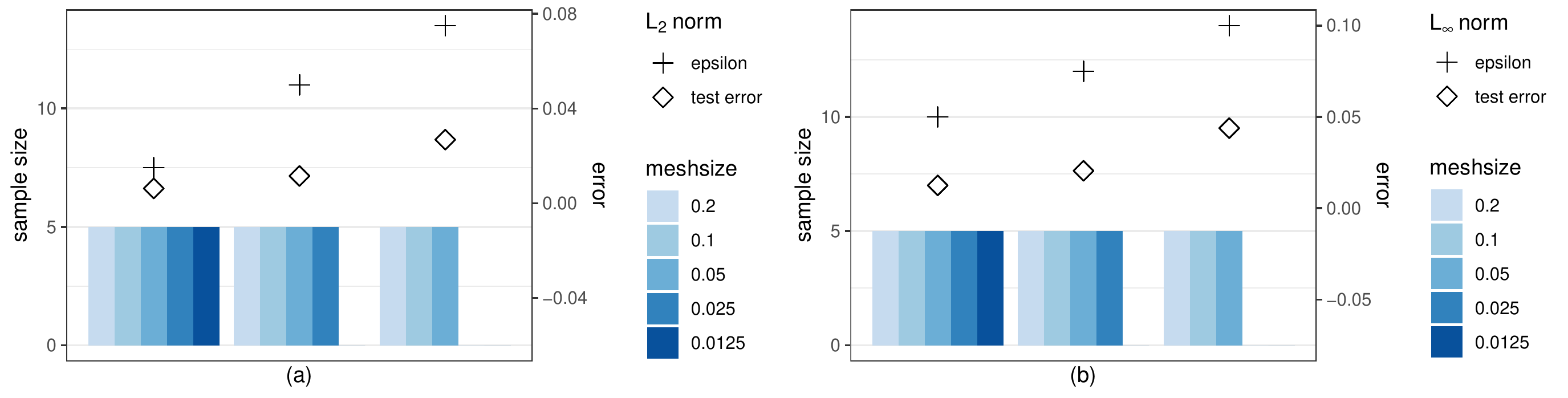}
    \caption{Visualizing the allocated sample sizes from stacking designs and the corresponding $L_2$ (left) and $L_\infty$ (right) test errors (marked $\diamondsuit$) at various error tolerances (marked $+$) for the Poisson's equation experiment.}
    \label{fig:poisson_epsilon}
\end{figure}

As before, we further explore the stacking designs for this problem with different error tolerances $\epsilon$ using both  $L_2$ and $L_{\infty}$ norms. Figure \ref{fig:poisson_epsilon} shows the sample sizes (at each fidelity level) and the corresponding errors of the final multi-level interpolator. We see again that the proposed designs can indeed consistently satisfy the desired error tolerance $\epsilon$. It is worth noting that, in this problem, a majority of the computational budget is expended on higher fidelity runs (i.e., with denser mesh densities), since the sample sizes are equally allocated over each fidelity level (see Figure \ref{fig:poisson_epsilon}). This is not too surprising: the fitted function is quite smooth ($\hat{\nu}=3.5$), and with estimated rate parameters $\hat{\alpha}\approx 1$ and $\hat{\beta}\approx 0.37$, the condition $\alpha/\beta < 2\nu/d$ can be shown to be satisfied (see earlier discussion of this condition in Section \ref{sec:complexitytheorem} and the cost complexity theorem). 

\subsection{Thermal Stress Analysis of Jet Engine Turbine Blade}\label{sec:casestudy2}
\begin{figure}[!t]
    \centering
    \includegraphics[width=\textwidth]{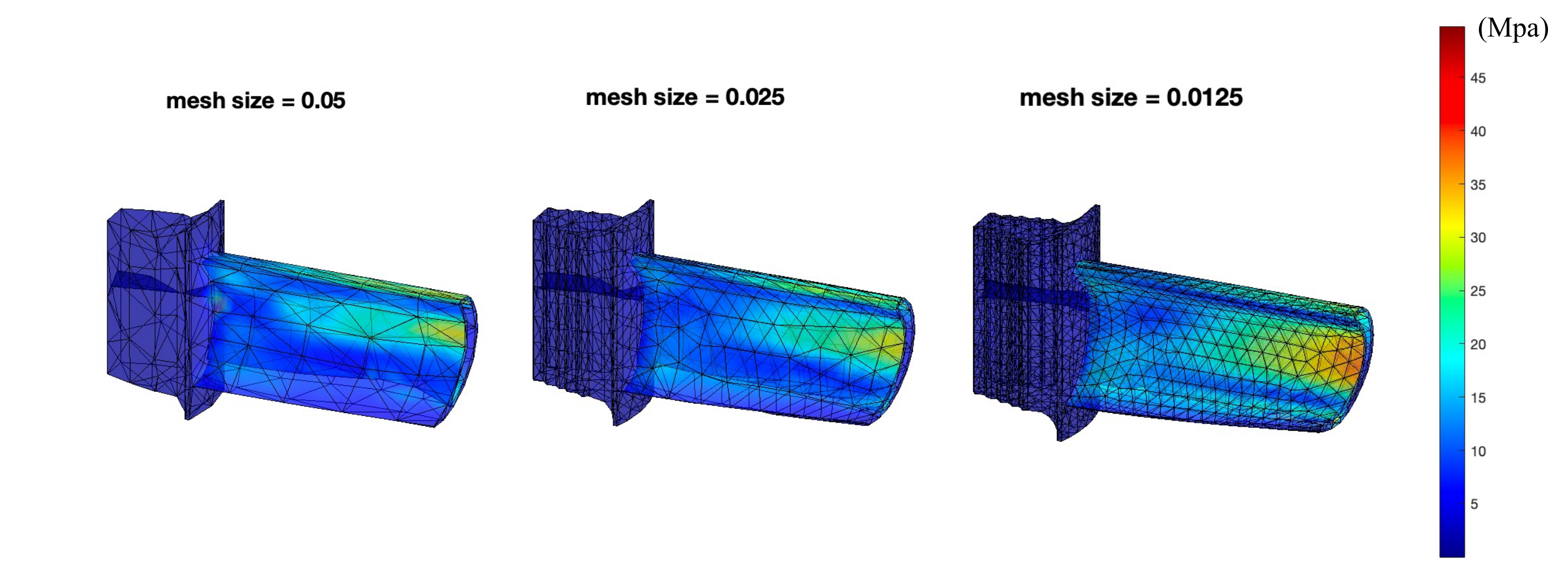}
    \caption{Visualizing the FEM solutions for three choices of mesh sizes in the turbine blade application.}
    \label{fig:mesh_blade}
\end{figure}

Finally, we investigate the performance of the proposed designs on a thermal stress analysis application for a jet turbine engine blade in steady-state operating condition. Here, the turbine is a component of the jet engine, composed of a radial array of blades typically made from nickel alloys that resist extremely high temperatures. To avoid mechanical failure and friction between the tip of the blade and the turbine casing, it is crucial that the blade design can withstand stress and deformations. See \cite{wright2006enhanced} and \cite{carter2005common} for more details. 
We thus wish to study here the effect of thermal stress and pressure of the surrounding gases on turbine blades. As before, this problem can be analyzed as a static structural model, which can be numerically solved using finite element modeling. The $d=2$ input variables are the pressure load on the pressure ($x_1$) and suction ($x_2$) sides of the blade, both of which range from 0.25 to 0.75 MPa, i.e., $x_1,x_2\in\Omega=[0.25,0.75]^2$. The response of interest is taken as the integral of the solution over the thermal stress profile. FEM simulations are performed using the Partial Differential Equation Toolbox in MATLAB \cite{MATLAB:R2021b}. Figure \ref{fig:mesh_blade} visualizes the blade structure and the simulated thermal stress profiles at three choices of mesh sizes. 

\begin{figure}[h]
    \centering
    \includegraphics[width=0.8\textwidth]{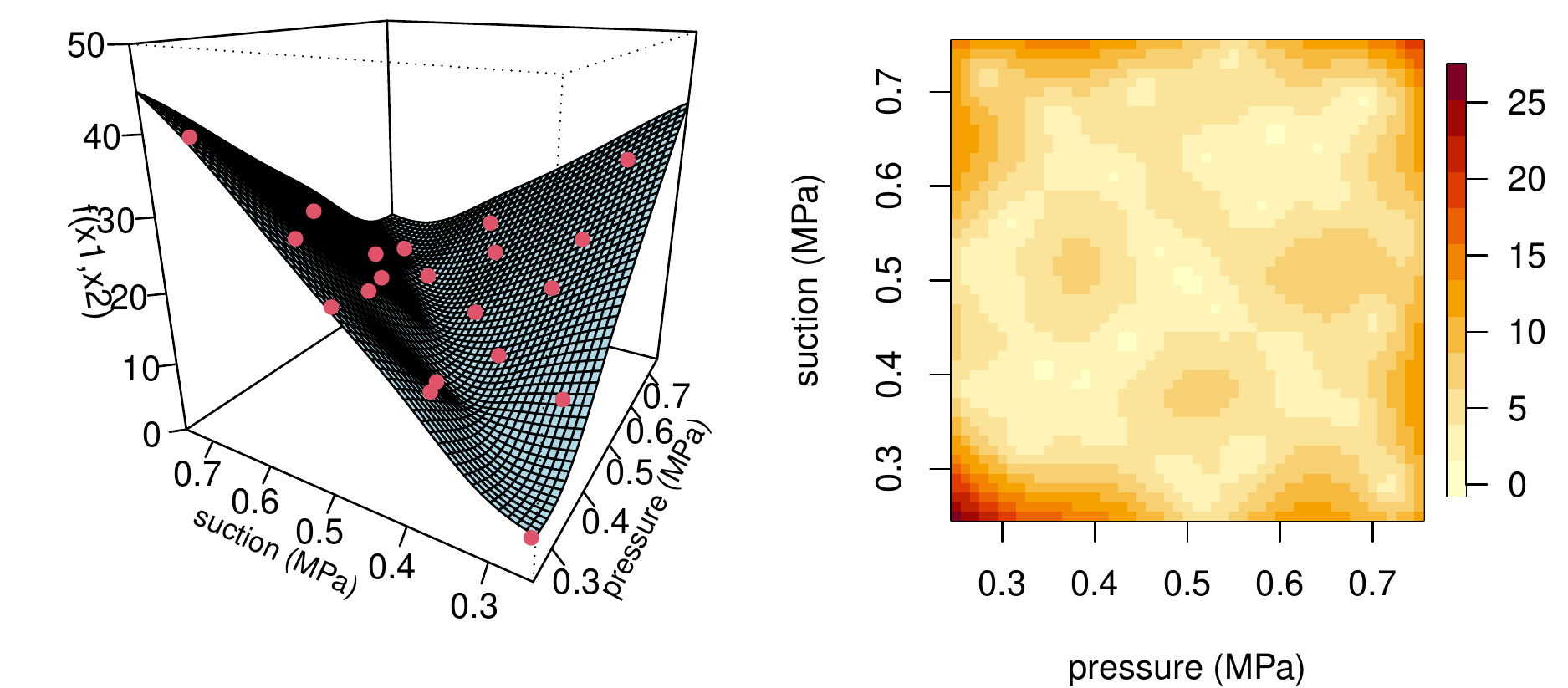}
    \caption{(Left) Visualizing the fitted multi-level interpolator with the test design points in red. (Right) Visualizing the pointwise error bounds \eqref{eq:predictiveinterval} over the design space.}
    \label{fig:blade_prediction}
\end{figure}

For stacking designs, we make use of a geometric sequence of fidelity parameters (here, mesh sizes) $\xi_l=\xi_0T^{-l}=0.1\times 2^{-l}$, $l = 1, 2, \cdots$. The desired prediction accuracy is then set to $\epsilon = 5$ in $L_2$-norm. Figure \ref{fig:blade_design} shows the proposed stacking designs over each iteration, Table \ref{tab:case_blade} summarizes the corresponding emulation and simulation error bounds. 
Here, we see that the multi-level interpolator requires $L=4$ iterations (resulting in $L=4$ fidelity levels for the final emulator) to achieve the desired prediction accuracy. Unlike in Section \ref{sec:casestudy}, the true function $f_{\infty}(x)$ cannot be expressed in closed form; we thus perform validation runs at 20 uniformly sampled input settings with a small mesh size of $\xi_5=3.125 \times 10^{-3}$. Figure  \ref{fig:blade_prediction} visualizes the final multi-level interpolator $\hat{f}_L(x_1,x_2)$ as in \eqref{eq:emu} with the out-of-sample test points (red points), as well as the pointwise error bounds \eqref{eq:predictiveinterval} over the input space. We see that the predicted response surface quite closely mimics the test data, which is as desired. This is confirmed by the empirical $L_2$-norm of prediction error on the test data (1.60), which is smaller than the desired error tolerance of $\epsilon=5$. This again shows that the proposed stacking designs can indeed achieve the desired prediction accuracy via iterative multi-fidelity modeling. Similar to previous subsections, we compare with the sequential design in \cite{le2015cokriging} with a similar experimental setup. The results indicate that, at the same computational cost, the $L_2$-error of the emulator using the sequential designs is 2.12 (with standard deviation of 0.83), which is higher than ours (which is  1.60).

\begin{table}[!t]
    \centering
    \begin{tabular}{ccccccc}
        \toprule
         & $L=1$  &  $L=2$ & $L=3$  & $L=4$ \\
        \midrule
        Mesh size & $\xi_1=0.05$ &$\xi_2=0.025$ & $\xi_3=0.0125$ & $\xi_4=0.00625$ \\
        Cost per run  (sec.) & $C_1=0.75$ &$C_2=1.07$ & $C_3=2.13$ & $C_4=11.51$ \\
        \midrule
        Bound of $\|f_{\infty}-f_{L}\|_{L_2(\Omega)}$ & NA & NA & 2.969 &  \textbf{0.956}\\
         &&&($\hat{\alpha}=0.81$)& ($\hat{\alpha}=1.09$)\\
      Bound of $\|f_L-\hat{f}_{L}\|_{L_2(\Omega)}$  & \textbf{2.324} & \textbf{2.408} &  \textbf{2.481} &  \textbf{2.491}\\
        \bottomrule
    \end{tabular}
    \caption{The estimated simulation and emulation error bounds (see \eqref{eq:stoppingrule} and \eqref{eq:powerfunctionboundL2}, respectively) at each design stage for the turbine blade application, with estimated rate parameter $\hat{\alpha}$ at stages $L=3$ and $L=4$. Bolded numbers indicate the error is less than $\epsilon/2$, where $\epsilon=5$ is the desired error tolerance. }
    \label{tab:case_blade}
\end{table}

\section{Concluding Remarks}\label{sec:discussion}
In this work, we proposed a novel stacking design approach for multi-fidelity modeling, which provides a sequential approach for designing multi-fidelity runs to achieve a desired prediction error $\epsilon > 0$. This addresses a key limitation of existing design methods, and allows for confident and cost-efficient predictive computing for a broad range of scientific and engineering problems. We then proved a cost complexity theorem which, under the employed multi-level RKHS interpolator, establishes a bound on computational cost (of training data simulation) needed to ensure a desired prediction error tolerance of $\epsilon$. A corollary then provides new insight on when the presented multi-level interpolator yields improved predictive performance over a single-fidelity RKHS interpolator. A suite of numerical experiments, including an application to jet turbine blade analysis, shows that the proposed method can efficiently and accurately emulate multi-fidelity computer experiments with some notion of confidence.

It is worth noting that alternative models, such as the one proposed by \cite{tuo2014surrogate}, offer interesting possibilities for addressing the challenges of multi-fidelity emulation. This model, which incorporates the rate at which the error with respect to the ideal/exact solution decreases, presents a different approach that could potentially enhance the efficiency of the emulation process. Future research can explore the utilization of this model to derive a similar error bound and develop a corresponding design methodology. In addition, we have identified recent studies, such as \cite{wang2020prediction,tuo2020kriging,wynne2021convergence}, which consider the impact of parameter misspecifications in the GPs and provide important extensions of stacking designs from a Bayesian perspective. In particular, these bounds can be used to investigate the emulation error in a GP assumption and determine  sample sizes under the constraints of a given computational budget (instead of a given  target error tolerance), as demonstrated in \cite{ehara2021}. Future research directions could involve 
exploring the optimal number of fidelity levels $L$  within a given computational budget. Finally, the development of potentially matching lower bounds for our cost-complexity theory would be useful in strengthening the insightful conclusions in Section \ref{sec:complexitytheorem}; we will investigate this as future work.

\vspace{0.5cm}
\noindent\textbf{Supplemental Materials}
Additional supporting materials can be found in Supplemental Materials, including the detailed proofs of Proposition \ref{thm:emulationaccuracy}, Theorem \ref{thm:complexity}, and Corollary \ref{cor:singlefidelity}, the detailed algorithm for Section \ref{forwardAlgo}, and the \texttt{R} code for reproducing the results in Section
\ref{sec:numericstudy}.

\bibliographystyle{siamplain}
\bibliography{bib}

\begin{thebibliography}{10}

\bibitem{ahlswede1987search}
{\sc R.~Ahlswede and I.~Wegener}, {\em Search Problems}, John Wiley \& Sons,
  Inc., 1987.

\bibitem{bect:hal-03096722}
{\sc J.~Bect, S.~Zio, G.~Perrin, C.~Cannamela, and E.~Vazquez}, {\em {On the
  quantification of discretization uncertainty: comparison of two paradigms}},
  in {14th World Congress in Computational Mechanics and ECCOMAS Congress 2020
  (WCCM-ECCOMAS)}, 2021.

\bibitem{breger2018points}
{\sc A.~Breger, M.~Ehler, and M.~Gr{\"a}f}, {\em Points on manifolds with
  asymptotically optimal covering radius}, Journal of Complexity, 48 (2018),
  pp.~1--14.

\bibitem{brenner2007fem}
{\sc S.~C. Brenner and L.~R. Scott}, {\em The Mathematical Theory of Finite
  Element Methods (Third Edition)}, New York: Springer., 2007.

\bibitem{caflisch1998monte}
{\sc R.~E. Caflisch}, {\em {M}onte {C}arlo and quasi-{M}onte {C}arlo methods},
  Acta Numerica, 7 (1998), pp.~1--49.

\bibitem{cao2021determining}
{\sc S.~Cao, Y.~Chen, J.~Coleman, J.~Mulligan, P.~Jacobs, R.~Soltz,
  A.~Angerami, R.~Arora, S.~Bass, L.~Cunqueiro, et~al.}, {\em Determining the
  jet transport coefficient $\hat{q}$ from inclusive hadron suppression
  measurements using {B}ayesian parameter estimation}, Physical Review C, 104
  (2021), p.~024905.

\bibitem{carter2005common}
{\sc T.~J. Carter}, {\em Common failures in gas turbine blades}, Engineering
  Failure Analysis, 12 (2005), pp.~237--247.

\bibitem{currin1988bayesian}
{\sc C.~Currin, T.~Mitchell, M.~Morris, and D.~Ylvisaker}, {\em A {B}ayesian
  approach to the design and analysis of computer experiments}, tech. report,
  Oak Ridge National Lab., TN (USA), 1988.

\bibitem{currin1991bayesian}
{\sc C.~Currin, T.~Mitchell, M.~Morris, and D.~Ylvisaker}, {\em Bayesian
  prediction of deterministic functions, with applications to the design and
  analysis of computer experiments}, Journal of the American Statistical
  Association, 86 (1991), pp.~953--963.

\bibitem{dick2013high}
{\sc J.~Dick, F.~Y. Kuo, and I.~H. Sloan}, {\em High-dimensional integration:
  the quasi-monte carlo way}, Acta Numerica, 22 (2013), pp.~133--288.

\bibitem{ehara2021}
{\sc A.~Ehara and S.~Guillas}, {\em An adaptive strategy for sequential designs
  of multilevel computer experiments}, International Journal for Uncertainty
  Quantification, 13 (2023), pp.~61--98.

\bibitem{evans2010partial}
{\sc L.~C. Evans}, {\em Partial Differential Equations (Second Edition)},
  vol.~19, American Mathematical Society, 2010.

\bibitem{everett2021multisystem}
{\sc D.~Everett, W.~Ke, J.-F. Paquet, G.~Vujanovic, S.~Bass, L.~Du, C.~Gale,
  M.~Heffernan, U.~Heinz, D.~Liyanage, et~al.}, {\em Multisystem {B}ayesian
  constraints on the transport coefficients of {QCD} matter}, Physical Review
  C, 103 (2021), p.~054904.

\bibitem{fang1993number}
{\sc K.-T. Fang and Y.~Wang}, {\em Number-Theoretic Methods in Statistics},
  vol.~51, CRC Press, 1993.

\bibitem{giles2008multilevel}
{\sc M.~B. Giles}, {\em Multilevel monte carlo path simulation}, Operations
  research, 56 (2008), pp.~607--617.

\bibitem{giles2015multilevel}
{\sc M.~B. Giles}, {\em Multilevel monte carlo methods}, Acta Numerica, 24
  (2015), pp.~259--328.

\bibitem{gramacy2020surrogates}
{\sc R.~B. Gramacy}, {\em Surrogates: Gaussian Process Modeling, Design, and
  Optimization for the Applied Sciences}, Chapman and Hall/CRC, 2020.

\bibitem{haaland2011accurate}
{\sc B.~Haaland and P.~Z.~G. Qian}, {\em Accurate emulators for large-scale
  computer experiments}, The Annals of Statistics, 39 (2011), pp.~2974--3002.

\bibitem{hundsdorfer2003numerical}
{\sc W.~H. Hundsdorfer and J.~G. Verwer}, {\em Numerical Solution of
  Time-dependent Advection-diffusion-reaction Equations}, vol.~33, Springer,
  2003.

\bibitem{ji2022graphical}
{\sc Y.~Ji, S.~Mak, D.~Soeder, J.-F. Paquet, and S.~A. Bass}, {\em A graphical
  multi-fidelity {G}aussian process model, with application to emulation of
  expensive computer simulations}, 2022,
  \url{https://arxiv.org/abs/2108.00306}.

\bibitem{ji2022multi}
{\sc Y.~Ji, H.~S. Yuchi, D.~Soeder, J.-F. Paquet, S.~A. Bass, V.~R. Joseph,
  C.~Wu, and S.~Mak}, {\em Multi-stage multi-fidelity {G}aussian process
  modeling, with application to heavy-ion collisions}, arXiv preprint
  arXiv:2209.13748,  (2022).

\bibitem{kanagawa2018gaussian}
{\sc M.~Kanagawa, P.~Hennig, D.~Sejdinovic, and B.~K. Sriperumbudur}, {\em
  Gaussian processes and kernel methods: A review on connections and
  equivalences}, arXiv preprint arXiv:1807.02582,  (2018).

\bibitem{kennedy2000predicting}
{\sc M.~C. Kennedy and A.~O'Hagan}, {\em Predicting the output from a complex
  computer code when fast approximations are available}, Biometrika, 87 (2000),
  pp.~1--13.

\bibitem{le2013bayesian}
{\sc L.~Le~Gratiet}, {\em Bayesian analysis of hierarchical multifidelity
  codes}, SIAM/ASA Journal on Uncertainty Quantification, 1 (2013),
  pp.~244--269.

\bibitem{le2015cokriging}
{\sc L.~Le~Gratiet and C.~Cannamela}, {\em Cokriging-based sequential design
  strategies using fast cross-validation techniques for multi-fidelity computer
  codes}, Technometrics, 57 (2015), pp.~418--427.

\bibitem{le2014recursive}
{\sc L.~Le~Gratiet and J.~Garnier}, {\em Recursive co-kriging model for design
  of computer experiments with multiple levels of fidelity}, International
  Journal for Uncertainty Quantification, 4 (2014).

\bibitem{lukic2001stochastic}
{\sc M.~Luki{\'c} and J.~Beder}, {\em Stochastic processes with sample paths in
  reproducing kernel hilbert spaces}, Transactions of the American Mathematical
  Society, 353 (2001), pp.~3945--3969.

\bibitem{mak2018minimax}
{\sc S.~Mak and V.~R. Joseph}, {\em Minimax and minimax projection designs
  using clustering}, Journal of Computational and Graphical Statistics, 27
  (2018), pp.~166--178.

\bibitem{mak2018support}
{\sc S.~Mak and V.~R. Joseph}, {\em Support points}, The Annals of Statistics,
  46 (2018), pp.~2562--2592.

\bibitem{mak2017efficient}
{\sc S.~Mak, C.-L. Sung, X.~Wang, S.-T. Yeh, Y.-H. Chang, V.~R. Joseph,
  V.~Yang, and C.~F.~J. Wu}, {\em An efficient surrogate model for emulation
  and physics extraction of large eddy simulations}, Journal of the American
  Statistical Association, 113 (2018), pp.~1443--1456.

\bibitem{MATLAB:R2021b}
{\sc MATLAB}, {\em {MATLAB version 9.11.0.1769968 (R2021b)}}, The Mathworks,
  Inc., Natick, Massachusetts, 2021.

\bibitem{muller2009komplexitat}
{\sc S.~M{\"u}ller}, {\em Komplexit{\"a}t und Stabilit{\"a}t von kernbasierten
  Rekonstruktionsmethoden}, PhD thesis, Nieders{\"a}chsische Staats-und
  Universit{\"a}tsbibliothek G{\"o}ttingen, 2009.

\bibitem{myren2021comparison}
{\sc S.~Myren and E.~Lawrence}, {\em A comparison of {G}aussian processes and
  neural networks for computer model emulation and calibration}, Statistical
  Analysis and Data Mining, 14 (2021), pp.~606--623.

\bibitem{oberkampf2010verification}
{\sc W.~L. Oberkampf and C.~J. Roy}, {\em Verification and Validation in
  Scientific Computing}, Cambridge university press, 2010.

\bibitem{Perdikaris_2017}
{\sc P.~Perdikaris, M.~Raissi, A.~Damianou, N.~D. Lawrence, and G.~E.
  Karniadakis}, {\em Nonlinear information fusion algorithms for data-efficient
  multi-fidelity modelling}, Proceedings of the Royal Society A: Mathematical,
  Physical and Engineering Sciences, 473 (2017), p.~20160751.

\bibitem{qian2008bayesian}
{\sc P.~Z.~G. Qian and C.~F.~J. Wu}, {\em Bayesian hierarchical modeling for
  integrating low-accuracy and high-accuracy experiments}, Technometrics, 50
  (2008), pp.~192--204.

\bibitem{richardson1911ix}
{\sc L.~F. Richardson}, {\em The approximate arithmetical solution by finite
  differences of physical problems involving differential equations, with an
  application to the stresses in a masonry dam}, Philosophical Transactions of
  the Royal Society of London. Series A, Containing Papers of a Mathematical or
  Physical Character, 210 (1911), pp.~307--357.

\bibitem{sacks1989design}
{\sc J.~Sacks, W.~J. Welch, T.~J. Mitchell, and H.~P. Wynn}, {\em Design and
  analysis of computer experiments}, Statistical Science, 4 (1989),
  pp.~409--423.

\bibitem{santner2003design}
{\sc T.~J. Santner, B.~J. Williams, and W.~I. Notz}, {\em The Design and
  Analysis of Computer Experiments (Second Edition)}, Springer New York, 2018.

\bibitem{scheuerer2013interpolation}
{\sc M.~Scheuerer, R.~Schaback, and M.~Schlather}, {\em Interpolation of
  spatial data--a stochastic or a deterministic problem?}, European Journal of
  Applied Mathematics, 24 (2013), pp.~601--629.

\bibitem{sobol1967distribution}
{\sc I.~M. Sobol'}, {\em On the distribution of points in a cube and the
  approximate evaluation of integrals}, Zhurnal Vychislitel'noi Matematiki i
  Matematicheskoi Fiziki, 7 (1967), pp.~784--802.

\bibitem{stein2012interpolation}
{\sc M.~L. Stein}, {\em Interpolation of Spatial Data: Some Theory for
  Kriging}, Springer Science \& Business Media, 2012.

\bibitem{stroh2022sequential}
{\sc R.~Stroh, J.~Bect, S.~Demeyer, N.~Fischer, D.~Marquis, and E.~Vazquez},
  {\em Sequential design of multi-fidelity computer experiments: maximizing the
  rate of stepwise uncertainty reduction}, Technometrics, 64 (2022),
  pp.~199--209.

\bibitem{teckentrup2020convergence}
{\sc A.~L. Teckentrup}, {\em Convergence of {G}aussian process regression with
  estimated hyper-parameters and applications in {B}ayesian inverse problems},
  SIAM/ASA Journal on Uncertainty Quantification, 8 (2020), pp.~1310--1337.

\bibitem{templeton2015calibration}
{\sc J.~A. Templeton, M.~L. Blaylock, S.~P. Domino, J.~C. Hewson, P.~R. Kumar,
  J.~Ling, H.~N. Najm, A.~Ruiz, C.~Safta, K.~Sargsyan, et~al.}, {\em
  Calibration and forward uncertainty propagation for large-eddy simulations of
  engineering flows}, tech. report, Sandia National Lab.(SNL-CA), Livermore, CA
  (United States); Sandia National Lab.(SNL-NM), Albuquerque, NM (United
  States), 2015.

\bibitem{tripathy2018deep}
{\sc R.~K. Tripathy and I.~Bilionis}, {\em {Deep UQ}: Learning deep neural
  network surrogate models for high dimensional uncertainty quantification},
  Journal of Computational Physics, 375 (2018), pp.~565--588.

\bibitem{tuo2020kriging}
{\sc R.~Tuo and W.~Wang}, {\em Kriging prediction with isotropic mat{\'e}rn
  correlations: Robustness and experimental designs}, The Journal of Machine
  Learning Research, 21 (2020), pp.~7604--7641.

\bibitem{tuo2020improved}
{\sc R.~Tuo, Y.~Wang, and C.~F.~J. Wu}, {\em On the improved rates of
  convergence for {M}at\`ern-type kernel ridge regression with application to
  calibration of computer models}, SIAM/ASA Journal on Uncertainty
  Quantification, 8 (2020), pp.~1522--1547.

\bibitem{tuo2014surrogate}
{\sc R.~Tuo, C.~F.~J. Wu, and D.~Yu}, {\em Surrogate modeling of computer
  experiments with different mesh densities}, Technometrics, 56 (2014),
  pp.~372--380.

\bibitem{wahba1990spline}
{\sc G.~Wahba}, {\em Spline Models for Observational Data}, SIAM, 1990.

\bibitem{wang2020prediction}
{\sc W.~Wang, R.~Tuo, and C.~F.~J. Wu}, {\em On prediction properties of
  kriging: Uniform error bounds and robustness}, Journal of the American
  Statistical Association, 115 (2020), pp.~920--930.

\bibitem{wendland2004scattered}
{\sc H.~Wendland}, {\em Scattered Data Approximation}, vol.~17, Cambridge
  university press, 2004.

\bibitem{wright2006enhanced}
{\sc L.~M. Wright and J.-C. Han}, {\em Enhanced internal cooling of turbine
  blades and vanes}, The Gas Turbine Handbook, 4 (2006), pp.~1--5.

\bibitem{wu1993local}
{\sc Z.-m. Wu and R.~Schaback}, {\em Local error estimates for radial basis
  function interpolation of scattered data}, IMA journal of Numerical Analysis,
  13 (1993), pp.~13--27.

\bibitem{wynne2021convergence}
{\sc G.~Wynne, F.-X. Briol, and M.~Girolami}, {\em Convergence guarantees for
  {G}aussian process means with misspecified likelihoods and smoothness},
  Journal of Machine Learning Research, 22 (2021), pp.~1--40.

\bibitem{xiu2010numerical}
{\sc D.~Xiu}, {\em Numerical methods for stochastic computations}, in Numerical
  Methods for Stochastic Computations, Princeton university press, 2010.

\bibitem{yakowitz2000global}
{\sc S.~Yakowitz, P.~L'ecuyer, and F.~Vazquez-Abad}, {\em Global stochastic
  optimization with low-dispersion point sets}, Operations Research, 48 (2000),
  pp.~939--950.

\end{thebibliography}

\newpage
\setcounter{page}{1}
\bigskip
\bigskip
\bigskip
\begin{center}
{\Large\bf Supplementary Materials for ``Stacking designs: designing multi-fidelity computer experiments with target predictive accuracy''}
\end{center}
\medskip

\setcounter{section}{0}
\setcounter{equation}{0}
\def\theequation{S\arabic{section}.\arabic{equation}}
\def\thesection{S\arabic{section}}

\section{Proof of Proposition \ref{thm:emulationaccuracy}}\label{append:theoremproofemulation}
Since $f_L=\sum^L_{l=1}(f_l-f_{l-1})$ and $\hat{f}_L=\sum^L_{l=1}\mathcal{P}_l$, it follows that 
\begin{align}\label{eq:theoremproof}
    |f_L(x)-\hat{f}_L(x)|&=\left|\sum^L_{l=1}(f_l(x)-f_{l-1}(x))-\sum^L_{l=1}\mathcal{P}_l(x)\right|\nonumber\\
    &=\left|\sum^L_{l=1}\left[(f_l(x)-f_{l-1}(x))-\mathcal{P}_l(x)\right]\right|\nonumber\\
    &\leq\sum^L_{l=1}\left|(f_l(x)-f_{l-1}(x))-\mathcal{P}_l(x)\right|,
\end{align}
where the last inequality follows the triangle inequality.

By Theorem 11.4 of \cite{wendland2004scattered}, it follows that
\begin{equation}\label{eq:theoremproof2}
    |(f_l(x)-f_{l-1}(x))-\mathcal{P}_l(x)|\leq\sigma_l(x)\|f_l-f_{l-1}\|_{\mathcal{N}_{\Phi_l}(\Omega)},
\end{equation}
where $\sigma_l(x)$ is the power function defined in \eqref{eq:powerfunction}. For a  Mat\'ern kernel  $\Phi_l$ given by \eqref{eq:maternkernel} with the smoothness parameter $\nu_l$, according to Lemma 2 of \cite{wang2020prediction}, which was derived from Theorem 5.14 of \cite{wu1993local}, there exist constants $c_l$ and $h_{l}$ such that 
\begin{equation}\label{eq:theoremproof3}
\sigma_l(x)\leq c_lh^{\nu_l}_{\mathcal{X}_l,\Theta_l}
\end{equation}
provided that $h_{\mathcal{X}_l,\Theta_l}\leq h_{l}$, where  $h_{\mathcal{X}_l,\Theta_l}:=\sup_{x\in\Omega}\min_{x_u\in \mathcal{X}_l}\|\Theta_l^{-1}(x-x_u)\|_2$. Since 
\begin{equation}\label{eq:theoremproof3}
\|\Theta_l^{-1}(x-x_u)\|_2\leq\|\Theta_l^{-1}\|_2\|x-x_u\|_2
\end{equation}
for any $x\in\Omega$ and $x_u\in\chi_l$, it can be shown that 
\begin{equation}\label{eq:theoremproof4}
h_{\mathcal{X}_l,\Theta_l}\leq \|\Theta_l^{-1}\|_2 h_{\mathcal{X}_l},
\end{equation}
where $h_{\mathcal{X}_l}=\sup_{x\in\Omega}\min_{x_u\in \mathcal{X}_l}\|x-x_u\|_2$.

Combining \eqref{eq:theoremproof}, \eqref{eq:theoremproof2}, \eqref{eq:theoremproof3}, and \eqref{eq:theoremproof4}, we have 
\begin{align*}
    |f_L(x)-\hat{f}_L(x)|\leq c_0\sum^L_{l=1}\|\Theta^{-1}_l\|^{\nu_l}_2h_{\mathcal{X}_l}^{\nu_l}\|f_l-f_{l-1}\|_{\mathcal{N}_{\Phi_l}(\Omega)},
\end{align*}
where $c_0=\max_{l=1,\ldots,L}c_l$, whose finiteness is ensured by the assumption that $\nu_l\in [\nu_{\min},\nu_{\max}]$, and the boundedness of $\|\Theta_l\|_2$ and $\|\Theta_l^{-1}\|_2$.

\section{Algorithm for multi-level interpolator with stacking design}\label{alg:stackingdesign}
\begin{algorithmic}[1]
\Require $\epsilon>0$
\Ensure $\|f_{\infty}-\hat{f}_L\|<\epsilon$
\State Set an initial sample $\mathcal{X}_0$ of size $n_0$. \Comment{$n_0\approx 5d \sim 10d$ is recommended}  
\State $L \gets 1$
\Repeat
\State Evaluate $f_L(x)$ on initial design $\mathcal{X}_0$
\State Estimate  parameters $\Theta_l$ and $\nu_l$ and  RKHS norm $\|f_l-f_{l-1}\|_{\mathcal{N}_{\Phi_{l}}(\Omega)}$
\State Set sample sizes $n_l$ via \eqref{eq:optss} and \eqref{eq:giventolerance}
\State Construct design $\mathcal{X}_l$ of size $n_l$ satisfying $\mathcal{X}_0\subseteq \mathcal{X}_L\subseteq \mathcal{X}_{L-1}\subseteq\cdots\subseteq \mathcal{X}_1$
\State Evaluate $f_l(x)$ on extra samples at each level as needed for new $\mathcal{X}_l$
\If{$L\geq 3$}
    \State Estimate rate parameter $\alpha$ via \eqref{eq:alphaestimate}
    \State Test for error convergence via \eqref{eq:stoppingrule}
\Else
    \State $L\gets L+1$
\EndIf
\Until converged via \eqref{eq:stoppingrule}
\State \Return multi-level interpolator via \eqref{eq:emu}
\end{algorithmic}

\section{Proof of Theorem \ref{thm:complexity}}\label{append:theoremproof}
Let $c_1=\sup_{x\in\Omega}c_1(x)$.
We start by choosing $L$ to be 
\begin{equation}\label{eq:L}
    L=\bigg\lceil{\frac{\log(2c_1\xi_0^{\alpha}\epsilon^{-1})}{\alpha\log T}}\bigg\rceil,
\end{equation}
implying 
$$\frac{\log(2c_1\xi_0^{\alpha}\epsilon^{-1})}{\alpha\log T}\leq L<\frac{\log(2c_1\xi_0^{\alpha}\epsilon^{-1})}{\alpha\log T}+1,$$
so that 
\begin{equation}\label{eq:condition}
    \frac{1}{2}T^{-\alpha}\epsilon<c_1\xi^{\alpha}_L\leq\frac{1}{2}\epsilon,
\end{equation}
and hence, by Condition 1,
\begin{equation}\label{eq:discrepancy}
    |f_{\infty}(x)-f_L(x)|\leq c_1(x)\xi^{\alpha}_L\leq c_1\xi^{\alpha}_L\leq\frac{1}{2}\epsilon.
\end{equation}
By Condition 2, it follows  that 
\begin{align}\label{eq:equivalent}
    \|f_l-f_{l-1}\|^2_{\mathcal{N}_{\Phi_l}(\Omega)}&=\langle f_l-f_{l-1},f_l-f_{l-1}\rangle_{\mathcal{N}_{\Phi_l}(\Omega)}\nonumber\\
    &=\langle f_l-f_{l-1},v_l\rangle_{L_2(\Omega)}\nonumber\\
    &\leq \|f_l-f_{l-1}\|_{L_2(\Omega)}\|v_l\|_{L_2(\Omega)}\nonumber\\   
    &\leq \bar{v}(\|f_\infty-f_l\|_{L_2(\Omega)}+\|f_\infty-f_{l-1}\|_{L_2(\Omega)})\nonumber\\     
    &\leq c_1\bar{v}{\rm{Vol}}(\Omega)(\xi^{\alpha}_l+\xi^{\alpha}_{l-1})\leq c_1\bar{v}{\rm{Vol}}(\Omega)(1+T^{\alpha})\xi^{\alpha}_{l},
\end{align}
where $\text{Vol}(\Omega)$ is the volume of $\Omega$. Then, combining \eqref{eq:equivalent} and Conditions 3 and 4 with Theorem \ref{thm:emulationaccuracy}, it follows that 
\begin{align}\label{eq:sobolevnorm}
    |f_L(x)-\hat{f}_{L}(x)|&\leq c_0\sum^L_{l=1}\|\Theta^{-1}_l\|^{\nu}_2h_{\mathcal{X}_l}^{\nu}\|f_l-f_{l-1}\|_{\mathcal{N}_{\Phi_l}(\Omega)}\nonumber\\
    &\leq c_0c^{1/2}_1c^{\nu}_2c^{\nu}_3\bar{v}^{1/2}{\rm{Vol}}(\Omega)^{1/2}(1+T^{\alpha})^{1/2}\sum^L_{l=1}n_l^{-\frac{\nu}{d}}\xi^{\alpha/2}_{l}.
\end{align}
Thus, by combining the equations \eqref{eq:discrepancy} and \eqref{eq:sobolevnorm}, we have 
\begin{align*}
    |f_{\infty}(x)-\hat{f}_{L}(x)|&\leq|f_{\infty}(x)-f_L(x)|+|f_L(x)-\hat{f}_{L}(x)|\\
    &\leq \frac{\epsilon}{2}+c_0c^{1/2}_1c^{\nu}_2c^{\nu}_3\bar{v}^{1/2}{\rm{Vol}}(\Omega)^{1/2}(1+T^{\alpha})^{1/2}\sum^L_{l=1}n_l^{-\frac{\nu}{d}}\xi^{\alpha/2}_{l}.
\end{align*}
The second term will be  discussed separately given $\alpha d=2\beta \nu,\alpha d>2\beta \nu$ and $\alpha d<2\beta \nu$. For notational simplicity, we let $c_6=c_0c^{1/2}_1c^{\nu}_2c^{\nu}_3\bar{v}^{1/2}{\rm{Vol}}(\Omega)^{1/2}(1+T^{\alpha})^{1/2}$.

If $\alpha d=2\beta \nu$, we set $n_l=\bigg\lceil{\left(2L\epsilon^{-1}c_6\right)^{\frac{d}{\nu}}}\xi^{\beta}_l\bigg\rceil$ so that 
$$
|f_{\infty}(x)-\hat{f}_L(x)|\leq \frac{\epsilon}{2}+c_6\sum^L_{l=1}n^{-\frac{\nu}{d}}_l\xi^{\alpha/2}_l\leq \frac{\epsilon}{2}+c_6\frac{\epsilon}{2Lc_6}L\xi^{\frac{\alpha}{2}-\frac{\nu\beta}{d}}_L\leq\frac{\epsilon}{2}+\frac{\epsilon}{2}\leq \epsilon.
$$
To bound the computational cost $C_{\rm tot}$, since the upper bound for $n_l$ is given by
$$
n_l\leq \left(2L\epsilon^{-1}c_6\right)^{\frac{d}{\nu}}\xi^{\beta}_l+1,
$$
the computational cost is bounded by
\begin{align}\label{eq:ctinequality}
    C_{\rm tot}\leq c_4\sum^L_{l=1}n_l\xi^{-\beta}_l&\leq c_4\left((2c_6)^{\frac{d}{\nu}}\epsilon^{-\frac{d}{\nu}}L^{\frac{d}{\nu}}\sum^L_{l=1}\xi^{\beta-\beta}_l+\sum^L_{l=1}\xi^{-\beta}_l\right)\nonumber\\&\leq c_4\left((2c_6)^{\frac{d}{\nu}}\epsilon^{-\frac{d}{\nu}}L^{1+\frac{d}{\nu}}+\sum^L_{l=1}\xi^{-\beta}_l\right).
\end{align}
By \eqref{eq:L}, the upper bound on $L$ is given by
$$
L\leq\frac{\log\epsilon^{-1}}{\alpha\log T}+\frac{\log(2c_1\xi_0^{\alpha})}{\alpha\log T}+1.
$$
Given that $1<\log \epsilon^{-1}$ for $\epsilon<e^{-1}$, it follows that 
\begin{equation}\label{eq:beta1L}
    L\leq c_7\log\epsilon^{-1},
\end{equation}
where 
$$
c_7=\frac{1}{\alpha\log T}+\max\left(0,\frac{\log(2c_1\xi_0^{\alpha})}{\alpha\log T}\right)+1.
$$
Moreover, since  $\epsilon^{-1/
\alpha}\leq\epsilon^{-\frac{d}{\beta \nu}}$ for $\alpha\geq \frac{\beta \nu}{d}$ and   $\epsilon<e^{-1}$, by \eqref{eq:condition} it follows
\begin{equation}\label{eq:xibound}
  \xi^{-1}_L<T\left(\frac{\epsilon}{2c_1}\right)^{-1/\alpha}<T2^{1/\alpha}c_1^{1/\alpha}\epsilon^{-\frac{d}{\beta \nu}}.    
\end{equation}
Then, by the standard result for a geometric series and the inequality in \eqref{eq:xibound},
\begin{align}\label{eq:xisumbound}
    \sum^L_{l=1}\xi_l^{-\beta}=\xi_L^{-\beta}\sum^L_{l=1}(T^{-\beta})^{l-L}&<\xi_L^{-\beta}(1-T^{-\beta})^{-1}\nonumber\\&<2^{\beta/\alpha}c_1^{\beta/\alpha}\frac{T^{\beta}}{1-T^{-\beta}}\epsilon^{-\frac{d}{\nu}}.
\end{align}
Thus, combining \eqref{eq:ctinequality}, \eqref{eq:beta1L}, and \eqref{eq:xisumbound}, and by the fact that $1<\log\epsilon^{-1}$ for $\epsilon<e^{-1}$, it follows that 
$$C_{\rm tot}\leq c_5\epsilon^{-\frac{d}{\nu}}\log(\epsilon^{-1})^{1+\frac{d}{\nu}},$$
where 
$
c_5=\left(c_42^{\frac{d}{\nu}}c_6^{\frac{d}{\nu}}c^{1+\frac{d}{\nu}}_7+c_42^{\frac{\beta}{\alpha}}c_1^{\frac{\beta}{\alpha}}\frac{T^{\beta}}{1-T^{-\beta}}\right).
$

If $\alpha d>2\beta \nu$, we set $n_l=\bigg\lceil\left(2\epsilon^{-1}c_6\xi_0^{\frac{\alpha d-2\beta \nu}{2(\nu+d)}}(1-T^{-\frac{\alpha d-2\beta \nu}{2(\nu+d)}})^{-1}\right)^{\frac{d}{\nu}}\xi^{\frac{(\alpha+2\beta)d}{2(\nu+d)}}_l\bigg\rceil$ so that 
\begin{align*}
    |f_{\infty}(x)-\hat{f}_L(x)|&\leq \frac{\epsilon}{2}+\frac{\epsilon}{2} \xi_0^{-\frac{\alpha d-2\beta \nu}{2(\nu+d)}}\left(1-T^{-\frac{\alpha d-2\beta \nu}{2(\nu+d)}}\right) \sum_{l=1}^{L} \xi_{l}^{\frac{\alpha}{2}-\frac{\nu}{d}\frac{(\alpha+2\beta)d}{2(\nu+d)}}\\
    &=\frac{\epsilon}{2}+\frac{\epsilon}{2} \xi_0^{-\frac{\alpha d-2\beta \nu}{2(\nu+d)}}\left(1-T^{-\frac{\alpha d-2\beta \nu}{2(\nu+d)}}\right) \sum_{l=1}^{L}\xi_0^{\frac{\alpha d-2\beta \nu}{2(\nu+d)}}T^{-\frac{\alpha d-2\beta \nu}{2(\nu+d)}l}\\
    &\leq\frac{\epsilon}{2}+\frac{\epsilon}{2} \left(1-T^{-\frac{\alpha d-2\beta \nu}{2(\nu+d)}}\right) \left(1-T^{-\frac{\alpha d-2\beta \nu}{2(\nu+d)}}\right)^{-1}=\epsilon.
\end{align*}
To bound the computational cost $C_{\rm tot}$, since the upper bound for $n_l$ is given by
$$
n_l\leq \left(2\epsilon^{-1}c_6\xi_0^{\frac{\alpha d-2\beta \nu}{2(\nu+d)}}(1-T^{-\frac{\alpha d-2\beta \nu}{2(\nu+d)}})^{-1}\right)^{\frac{d}{\nu}}\xi^{\frac{(\alpha+2\beta)d}{2(\nu+d)}}_l+1,
$$
the computational cost is bounded by
\begin{equation}\label{eq:ctinequality2}
    C_{\rm tot}\leq c_4\sum^L_{l=1}n_l\xi^{-\beta}_l\leq c_4\left(c_8\epsilon^{-\frac{d}{\nu}}\sum^L_{l=1}\xi^{\frac{(\alpha+2\beta)d}{2(\nu+d)}-\beta}_l+\sum^L_{l=1}\xi^{-\beta}_l\right),
\end{equation}
where $c_8=(2c_6\xi_0^{\frac{\alpha d-2\beta \nu}{2(\nu+d)}}(1-T^{-\frac{\alpha d-2\beta \nu}{2(\nu+d)}})^{-1})^{\frac{d}{\nu}}$.  By the standard result for a geometric series, we have
\begin{align}\label{eq:xisumbound2}
    \sum^L_{l=1}\xi^{\frac{(\alpha+2\beta)d}{2(\nu+d)}-\beta}_l=\sum^L_{l=1}\xi^{\frac{\alpha d-2\beta \nu}{2(\nu+d)}}_l=\xi_0^{\frac{\alpha d-2\beta \nu}{2(\nu+d)}}\sum^L_{l=1}T^{-\frac{\alpha d-2\beta \nu}{2(\nu+d)}l}\leq \xi_0^{\frac{\alpha d-2\beta \nu}{2(\nu+d)}}\left(1-T^{-\frac{\alpha d-2\beta \nu}{2(\nu+d)}}\right)^{-1}.
\end{align}
Thus, combining \eqref{eq:ctinequality2}, \eqref{eq:xisumbound2} and \eqref{eq:xisumbound}, it follows that 
$$C_{\rm tot}\leq c_5\epsilon^{-\frac{d}{\nu}},$$
where 
$$
c_5=c_4c_8\xi_0^{\frac{\alpha d-2\beta \nu}{2(\nu+d)}}\left(1-T^{-\frac{\alpha d-2\beta \nu}{2(\nu+d)}}\right)^{-1}+c_42^{\frac{\beta}{\alpha}}c_1^{\frac{\beta}{\alpha}}\frac{T^{\beta}}{1-T^{-\beta}}.
$$

If $\alpha d<2\beta \nu$, we set $n_l=\bigg\lceil\left(2\epsilon^{-1}c_6\xi_L^{\frac{\alpha d-2\beta \nu}{2(\nu+d)}}(1-T^{-\frac{2\beta\nu -\alpha d}{2(\nu+d)}})^{-1}\right)^{\frac{d}{\nu}}\xi^{\frac{(\alpha+2\beta)d}{2(\nu+d)}}_l\bigg\rceil$. Because  
\begin{align}\label{eq:xi3}
    \sum^L_{l=1}\xi_l^{-\frac{2\beta\nu -\alpha d}{2(\nu+d)}}=\xi_L^{-\frac{2\beta \nu-\alpha d}{2(\nu+d)}}\sum^L_{l=1}(T^{-\frac{2\beta \nu-\alpha d}{2(\nu+d)}})^{l-L}<\xi_L^{-\frac{2\beta \nu-\alpha d}{2(\nu+d)}}(1-T^{-\frac{2\beta \nu-\alpha d}{2(\nu+d)}})^{-1},
\end{align}
it follows that 
\begin{align*}
    |f_{\infty}(x)-\hat{f}_L(x)|&\leq \frac{\epsilon}{2}+\frac{\epsilon}{2} \xi_L^{-\frac{\alpha d-2\beta \nu}{2(\nu+d)}}(1-T^{-\frac{2\beta\nu-\alpha d}{2(\nu+d)}}) \sum_{l=1}^{L} \xi_{l}^{\frac{\alpha}{2}-\frac{\nu}{d}\frac{(\alpha+2\beta)d}{2(\nu+d)}}\\
    &= \frac{\epsilon}{2}+\frac{\epsilon}{2} \xi_L^{-\frac{\alpha d-2\beta \nu}{2(\nu+d)}}(1-T^{-\frac{2\beta \nu-\alpha d}{2(\nu+d)}}) \sum_{l=1}^{L} \xi_{l}^{-\frac{2\beta \nu-\alpha d}{2(\nu+d)}}\\
    &\leq\frac{\epsilon}{2}+\frac{\epsilon}{2}=\epsilon.
\end{align*}
To bound the computational cost $C_{\rm tot}$, since the upper bound for $n_l$ is given by
$$
n_l\leq \left(2\epsilon^{-1}c_6\xi_L^{\frac{\alpha d-2\beta \nu}{2(\nu+d)}}(1-T^{-\frac{2\beta \nu-\alpha d}{2(\nu+d)}})^{-1}\right)^{\frac{d}{\nu}}\xi^{\frac{(\alpha+2\beta)d}{2(\nu+d)}}_l+1,
$$
the computational cost is bounded by
\begin{equation}\label{eq:ctinequality3}
    C_{\rm tot}\leq c_4\sum^L_{l=1}n_l\xi^{-\beta}_l\leq c_4\left(c_9\epsilon^{-\frac{d}{\nu}}\sum^L_{l=1}\xi^{\frac{(\alpha+2\beta)d}{2(\nu+d)}-\beta}_l+\sum^L_{l=1}\xi^{-\beta}_l\right),
\end{equation}
where $c_9=(2c_6\xi_L^{\frac{\alpha d-2\beta \nu}{2(\nu+d)}}(1-T^{-\frac{2\beta \nu-\alpha d}{2(\nu+d)}})^{-1})^{\frac{d}{\nu}}$. 
Because \eqref{eq:condition} gives
\begin{equation*}
  \xi^{-1}_L<T\left(\frac{\epsilon}{2c_1}\right)^{-1/\alpha}<T2^{1/\alpha}c_1^{1/\alpha}\epsilon^{-1/\alpha},    
\end{equation*}
combining with  \eqref{eq:xi3}, it follows that 
\begin{align}\label{eq:xisumbound3}
    \sum^L_{l=1}\xi^{\frac{(\alpha+2\beta)d}{2(\nu+d)}-\beta}_l=\sum^L_{l=1}\xi^{\frac{\alpha d-2\beta \nu}{2(\nu+d)}}_l\leq\xi_L^{-\frac{2\beta \nu-\alpha d}{2(\nu+d)}}(1-T^{-\frac{2\beta \nu-\alpha d}{2(\nu+d)}})^{-1}\leq c_{10}\epsilon^{-\frac{2\beta \nu-\alpha d}{2\alpha(\nu+d)}},
\end{align}
where $$c_{10}=T^{\frac{2\beta \nu-\alpha d}{2(\nu+d)}}2^{\frac{2\beta \nu-\alpha d}{2\alpha(\nu+d)}}c_1^{\frac{2\beta \nu-\alpha d}{2\alpha(\nu+d)}}(1-T^{-\frac{2\beta \nu-\alpha d}{2(\nu+d)}})^{-1}.$$
Thus, combining \eqref{eq:ctinequality3}, \eqref{eq:xisumbound3}, and \eqref{eq:xisumbound}, and by the fact $\epsilon^{-\frac{d}{\nu}}<\epsilon^{-\frac{d}{\nu}-\frac{2\beta\nu-\alpha d}{2\alpha(\nu+d)}}$ for $\epsilon<e^{-1}$, it follows that 
$$C_{\rm tot}\leq c_5\epsilon^{-\frac{d}{\nu}-\frac{2\beta \nu-\alpha d}{2\alpha(\nu+d)}},$$
where 
$$
c_5=c_4c_9c_{10}+c_42^{\frac{\beta}{\alpha}}c_1^{\frac{\beta}{\alpha}}\frac{T^{\beta}}{1-T^{-\beta}}.
$$

\section{Proof of Corollary \ref{cor:singlefidelity}}\label{append:corollaryproof}
By Condition 1 of Theorem \ref{thm:complexity} and Condition 1, it follows that  
\begin{equation}\label{eq:gbias}
    |f_{\infty}(x)-f_H(x)|\leq c_1(x)\xi^{\alpha}_H\leq c_1\xi^{\alpha}_H\leq\epsilon/2.
\end{equation}
Similar to Theorem \ref{thm:emulationaccuracy}, \begin{equation}\label{eq:gvar}
|f_H(x)-\hat{f}_H(x)|\leq c_{H}\|\Theta^{-1}_{H}\|^{\nu}_2h_{\mathcal{X}_H}^{\nu}\|f_H\|_{\mathcal{N}_{\Phi_H}(\Omega)}
\end{equation}
with a positive constant $c_{H}$. Given Condition 2, similar to \eqref{eq:equivalent}, it can be shown that 
$\|f_H\|_{\mathcal{N}_{\Phi_H}(\Omega)}\leq c_{11}\xi_H^{\alpha/2}$ with a positive constant $c_{11}$. Combining with \eqref{eq:gbias}, \eqref{eq:gvar}, by Condition 4 of Theorem \ref{thm:complexity} and Condition 3,  it follows
\begin{align*}
    |f_{\infty}(x)-\hat{f}_{H}(x)|&\leq|f_{\infty}(x)-f_H(x)|+|f_H(x)-\hat{f}_{H}(x)|\leq \frac{\epsilon}{2}+c_{12}n_H^{-\frac{\nu}{d}}\xi^{\alpha/2}_{H}
\end{align*}
with a positive constant $c_{12}$. Let $n_H=\bigg\lceil{\left(2\epsilon^{-1}c_{12}\xi^{\alpha/2}\right)^{\frac{d}{\nu}}}\bigg\rceil$, then it can be shown that $c_{12}n_H^{-\frac{\nu}{d}}\xi^{\alpha/2}_{l}\leq\epsilon/2$ which leads to $|f_{\infty}(x)-\hat{f}_{H}(x)|\leq\epsilon$. 

The bound the computational cost $C_{\rm{tot}}$, since the upper bound for $n_H$ is given by 
$$
n_H\leq\left(2\epsilon^{-1}c_{12}\xi^{\alpha/2}\right)^{\frac{d}{\nu}}+1,
$$
the computational cost is bounded by 
\begin{equation}\label{eq:gtotalcost}
    C_{\rm tot}\leq c_4n_H\xi^{-\beta}_H\leq c_{13}\epsilon^{-\frac{d}{\nu}}\xi_H^{\frac{\alpha d}{2\nu}-\beta}+\xi^{-\beta}_H,
\end{equation}
where $c_{13}=(2c_{12})^{d/\nu}$. By Condition 1, it follows that 
$$
\xi_H^{\frac{\alpha d}{2\nu}-\beta}\leq \left(\frac{\epsilon}{2c_1}\right)^{\frac{d}{2\nu}-\frac{\beta}{\alpha}}
$$
and 
$$
\xi_H^{-\beta}\leq c_1^{-\frac{\beta}{\alpha}}\left(\frac{\epsilon}{2}\right)^{-\frac{\beta}{\alpha}-\frac{ d}{2\nu}},
$$
and combining the two inequalities with \eqref{eq:gtotalcost}, we have 
$$
C_{\rm tot}\leq c_{13}\epsilon^{-\frac{d}{\nu}}\xi_H^{\frac{\alpha d}{2\nu}-\beta}+\xi^{-\beta}_H\leq c_{14}\epsilon^{-\frac{d}{\nu}+\frac{d}{2\nu}-\frac{\beta}{\alpha}}+c_{15}\epsilon^{-\frac{\beta}{\alpha}-\frac{ d}{2\nu}}\leq c_{16}\epsilon^{-\frac{\beta}{\alpha}-\frac{ d}{2\nu}},
$$
where $c_{16}=c_{14}+c_{15},c_{14}=c_{13}(2c_1)^{\beta/\alpha-d/(2\nu)}$, and $c_{15}=c_{13}c^{-\beta/\alpha}_12^{d/(2\nu)-\beta/\alpha}$. This finishes the proof.

\end{document}